\documentclass{iopjournal}
\usepackage[square,sort,comma,numbers]{natbib}
\usepackage{url}
\usepackage{graphicx,color} 
\usepackage{hyphenat} 
\usepackage{amsmath, amssymb, amsfonts,amsbsy,amsthm} 
\usepackage{float}
\usepackage{subfig}
\usepackage{soul,xcolor} 
\usepackage{tikz}
\usepackage{pgfpages}
\usepackage{tikz-3dplot}
\usepackage{hyperref}
\usepackage{tabularx}
\usepackage{multirow}
\usepackage{booktabs}
\usepackage{siunitx}
\usepackage{lineno}
\usepackage{amsfonts}
\usepackage{bm}
\usepackage{array}
\usepackage{soul}

\begin{document}

\articletype{Topical Review}

\title{Environmental noise challenges for the Einstein Telescope - A Review}

\author{Matteo Di Giovanni$^{1,2*}$\orcid{0000-0003-4049-8336}}

\affil{$^1$Scuola Normale Superiore, Pisa, I-56126}

\affil{$^2$Sezione di Pisa, Istituto Nazionale di Fisica Nucleare, Pisa, I-56126}

\affil{$^*$Author to whom any correspondence should be addressed.}

\email{matteo.digiovanni@sns.it}


\begin{abstract}
Einstein Telescope (ET) will be a next-generation ground based interferometric gravitational-wave detector designed to explore the Universe to an unprecedented depth and detail. With its 10–15 km long arms, underground infrastructure and advanced cryogenic technologies to suppress thermal noise, ET aims at achieving a significant leap forward in sensitivity with respect to current detectors such as Virgo, LIGO, and KAGRA. The coverage of the low-frequency band, starting at 2 Hz, will enable the observation of earlier inspiral phases of compact binary coalescences, enhancing early-warning capabilities and improving multi-messenger follow-up.
However, this increased sensitivity also implies a higher susceptibility to environmental disturbances, particularly in the 2–10 Hz range where anthropogenic, seismic and infrastructural noise noise are most relevant. Therefore, understanding and mitigating environmental contributions are essential to fully exploit the scientific potential of ET.
This chapter reviews the ongoing studies on environmental noise for ET, summarizes the main challenges related to site characterization and infrastructure design and outlines the strategies currently under development to ensure optimal detector performance.
\end{abstract}

\section{Introduction}
 \label{sec:Introduction}

The decade between 2015 and 2025 marked an unprecedented revolution in physics and astronomy. The first detection of gravitational waves (GW) from a binary black hole (BBH) merger \citep{GW150914} changed the perspective with which we observe the Universe. For the first time, physicists and astronomers were able to observe an event inaccessible to observatories that rely on the emission of photons from the source and to test the validity of the General Relativity theory. Moreover, the successful exploitation of ten years of observations \citep{gwtc1, gwtc2, GWTC3, gwtc4, gwtc5} led to the discovery of over 300 confirmed GW events, all from the merger of compact objects (BBH, binary neutron stars and neutron star black hole binaries), with increasing level of detail \cite{hawking, exploring}, thanks to the constant improvement of the sensitivity of current detectors, namely Advanced Virgo (AdVirgo), located in Italy \citep{aVirgo}, Advanced LIGO Hanford and Livingston (aLIGO), located in the United States \citep{aLIGO} and KAGRA in Japan \citep{kagra}. These detectors exploit the same basic working principle of a Michelson interferometer, but are uniquely designed to measure the tiny  path length differential variations that GW induce between the mirrors. This is achieved thanks to the inter-connection of complex sub-systems that benefit from state of the art technologies (e.g., advanced isolation systems for the mirrors \cite{superatt} or innovative materials for the mirror coatings and suspensions \cite{coating6}) to measure a path length variation smaller than the nucleus of an Hydrogen atom.

\begin{figure}[H]
\centering
\includegraphics[width=8.0 cm]{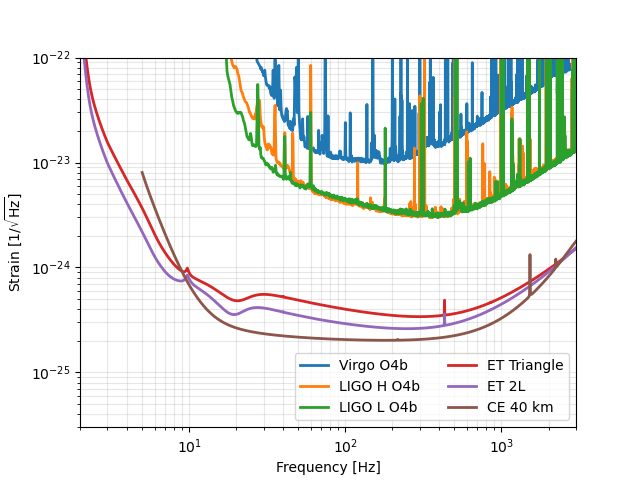}
\caption{Comparison of the reference sensitivities achieved during the O4b observing run by the LIGO and Virgo detectors \cite{o4bdatarelease} against the reference sensitivity of the ET detector in its triangular and 2L configurations \cite{etsenses} and of the 40 km CE detector \cite{cesenses}. These  curves represent a snapshot of a work-in-progress as of January 2026 (ET) and October 2024 (CE). They are not meant to reflect the final sensitivity of the detector.}\label{fig:LVKETsens}
\end{figure}

After the successful outcome of current detectors, the scientific community started to thoroughly investigate \cite{Maggiore_2020, coba} the prospects of future next generation GW interferometric detectors, to access increasingly large portions of the Universe and sources which are out of reach for current detectors. This new generation of detectors, i.e., Cosmic Explorer (CE) \cite{CE1, CE2, CE3} and the Einstein Telescope (ET) \cite{et, ET2010, ET2011, ET2020}, with the latter being the subject of this work, is expected to start observations in the late 2030s. ET was first proposed in 2010 and the foreseen improvements with respect to current detectors include the extension of the observation bandwidth from the current limit of about 20\,Hz to 2\,Hz, an improvement of the sensitivity up to two orders of magnitude at 20 Hz and a sensitivity gain of approximately a factor 8 beyond 100 Hz (Figure \ref{fig:LVKETsens}) \citep{ET2020}. 

Among all proposals for future ground-based detectors, ET is the most ambitious and is expected to be an innovation driver in terms of the development of the technologies needed to achieve the foreseen sensitivity. In fact, the expected improvement in sensitivity will not be solely due to the sheer increase in size of the detector, but also to the development and adoption of cutting-edge technologies which allow to reduce the contribution of the intrinsic noise levels of each subsystem of the detector. A first notable example, currently being tested in LIGO, Virgo and KAGRA, is quantum squeezing \cite{squeezed, squeezing1, squeezing2, squeezing3} to lessen the fluctuations in the number of photons reaching the photo-diode and the tiny variations in the mirrors’ positions due to the pressure exerted by fluctuating light itself. 

\begin{figure}[H]
\centering
\includegraphics[width=7.0 cm]{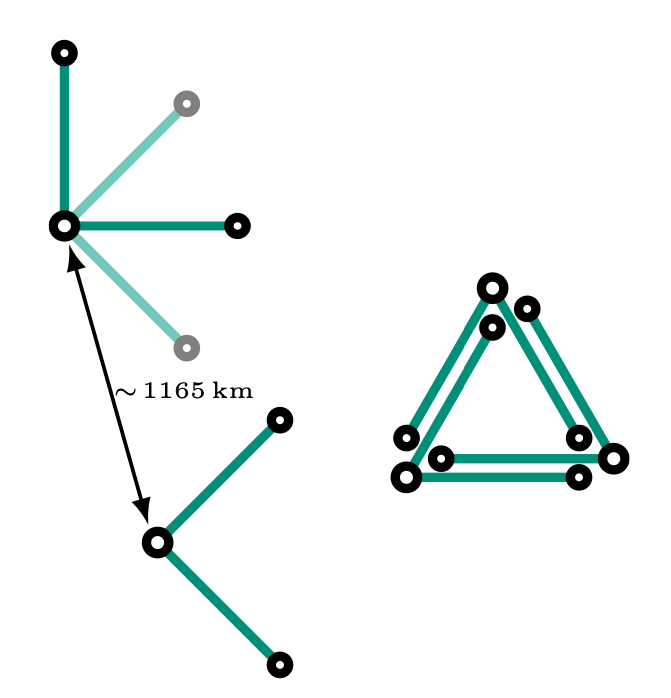}
\caption{A schematic picture of the different geometries considered for ET: two widely separated L-shaped detectors (either parallel or at $45^{\circ}$), or a triangle made of three nested detectors. This figure is taken from \cite{coba}. \label{xylophone}}
\end{figure}

For what concerns the configuration of ET, there are two proposals under consideration (Figure \ref{xylophone}). The most recent one is that of a detector network composed of two widely separated L-shaped pairs of detectors with 15 km long arms \cite{coba, Iacovelli_2024}. On the other hand, the original project foresees three pairs of nested interferometers arranged in an equilateral triangle with the sides 10\,km long \cite{et, ET2010, ET2011, ET2020}. Part of the innovation resides in the fact that, for each interferometer pair, one detector is optimized for low frequencies (LF, 2\,Hz $\lesssim f \lesssim$ 40\,Hz) and the other for high frequencies (HF, $f>$ 40\,Hz). A final comparative report on the 2L and triangular configurations is expected by the end of 2026. This document will be evaluated by the Board of Government Representatives\footnote{The Board of Government Representatives unites the governments who actively support the construction of the Einstein Telescope in Europe and forms a forum that links the involved governments and the funding agencies} of the ET collaboration to take the final decision. However, it is worth pointing out that some studies, notably Ref. \cite{Iacovelli_2024} plus others still in preparation, also investigate the potential of alternative configurations for ET, such as a single L-shaped detector combined with a triangular one.

Another remarkable example of the innovative technologies that are currently under development for ET originates from the need to operate a dedicated LF detector. In fact the HF sensitivity target requires high laser power, which translates into a high heat transfer to the mirrors, hence generating high thermal noise which is not compatible with the LF sensitivity target \cite{et, ET2010, ET2011, ET2020}. Accordingly, the choice was done to have a detector entirely dedicated to the HF range, with 3 MW of laser power circulating into the arms of the HF detector, whereas a second detector with 18 kW of laser power circulating in the arms will be dedicated to the LF range \cite{ET2020}. However, this is not enough to reach the LF sensitivity target. Thermal noise from the mirrors and suspensions at room temperature will still significantly limit the sensitivity of the detector. Therefore, the mirrors will be cooled down at a temperature between $\sim 10 \rm\, K$ and $\sim 20 \rm\, K$ with liquid Helium II \cite{Grohmann2021, koroveshi}. If we compare this requirement with current detectors, It is clearly understood that this is a complex task requiring substantial R$\&$D efforts, not only to define the baseline design of the cryogenic payloads (i.e. the unit composed by the mirror, the marionette, which coordinates the position of the mirror via suspension fibers, and the platform, from which the marionette is suspended and which, in its turn, is suspended from the room-temperature isolation system), responsible for heat extraction from the mirrors, but also to plan the infrastructure supporting the cryogenic operations. As a matter of comparison, the mirrors in LIGO and Virgo weight 40 kg, have a diameter of 35 cm and a thickness of 20 cm; in KAGRA, which operates at cryogenic temperatures, the mirrors are significantly smaller, being half in both weight and size. By contrast, the mirrors in ET-LF will weight 220 kg and have a diameter of 60 cm. Consequently, even though KAGRA provides relevant experience on the operations of an underground cryogenic GW detector, technical solutions cannot be copied 1:1 due to the downright difference in size of the detector's components and the different LF sensitivity target. It is also worth pointing out that Ref. \cite{dupletsa2026} investigated the impact of instrumental requirements on the scientific performance of ET and inferred that hypothetical issues in the reaching of the LF sensitivity target will not hinder the detection of astrophysical sources and a minimum detection threshold can still be achieved, albeit reducing the rate of detected sources.

A comprehensive overview of the state of the art in mirror production and in the baseline design of the cryogenic payloads of ET-LF is out of the scope of this review. Suffice it to say that one of the most promising solutions to effectively extract heat from such massive mirrors is to adopt monocrystalline silicon mirrors, preferred for their high thermal conductivity and low thermal expansion at low temperatures with respect to the sapphire mirrors adopted in KAGRA. In terms of payload design, at least two possible choices, which satisfy the LF sensitivity target, have been defined to extract heat from the mirrors: monocrystalline marionette suspension made of silicon or sapphire or a suspension tube made of titanium and filled with liquid Helium II \cite{koroveshi}. Mirror coatings also play a crucial role to reduce thermal noise in the cryogenic regimes without introducing either light absorption or scattering \cite{coating6}. A primary option, compatible with the foreseen laser wavelengths, is to use amorphous silicon multilayers \cite{coating1, coating4}. In parallel, research is exploring cutting-edge solutions like epitaxial crystalline coatings (colloquially known as ALGAS) and alternative crystalline nitrides or oxides \cite{coating5}, which can reduce mechanical losses albeit presenting complex fabrication challenges.

The LF sensitivity target also implies that the dedicated detector will be extremely susceptible to environmental noise sources, which dominate in that frequency band. As a consequence, accurate site selection is paramount to grant an adequately stable and quiet noise environment. In addition to that, the adoption of cryogenic technologies \cite{Puppo2011} for the mirrors is a technological challenge on its own: effectively cooling down 200 kg mirrors and their payload, without generating excessive environmental noise due to the operation of the cryogenic system, is a problem that is requiring constant commitment and dedication from the ET community. Moreover, in both the 2L and the triangle configurations, ET will be hosted underground, at a currently planned depth between 200\,m and 300\,m. The call for an underground detector originates from the nature of the noise sources that ET is expected to experience \cite{amann}. Special attention is given to terrestrial gravity perturbations \cite{harms}, i.e., fluctuations in the Earth's gravitational field caused by fluctuating seismic fields or atmospheric density fluctuations. These fluctuations, referred to as Newtonian Noise (NN), can limit the sensitivity of next generation ground based detectors and require complicated mitigation methods. However, the nature of the seismic field underground is different from that on the surface. Surface seismic waves decay exponentially with depth, leaving an underground dominant component formed by seismic body waves. This reduces the influence of NN from fluctuating seismic fields and has an effect on the complexity of mitigation methods. On top of that, another benefit of going underground can be found in the reduction of atmospheric disturbances in a controlled environment \cite{hutt} and from the fact that NN from atmospheric density fluctuations from infrasounds is greatly reduced underground \cite{mariusz}. Finally, underground seismic motion at the input of the suspension systems will also be reduced due to the decay of surface seismic waves with depth \cite{potsdam}. Consequently, NN mitigation to achieve the target sensitivity below 10 Hz is the main motivation to construct ET underground.

At this point, it is also worth pointing out the differences occurring between ET and CE in the overall approach to the design of a future generation ground based GW detector. In fact, CE foresees the construction of a network of two surface built widely separated L-shaped interferometers, one with 40 km long arms and the other with 20 km long arms \cite{CE1, CE2, CE3, CE4, nsfreport}, although alternate scenarios in which CE consists of a single 40 km facility, two 40 km facilities, or two 20 km facilities are under evaluation. The baseline configuration will be widely based on LIGO technology, including 1064 nm wavelength laser and, contrary to ET, room temperature operations. Other options are also under consideration
for future upgrades. Being of surface, it is also expected to adopt appropriate noise suppression and isolation systems to reach the design sensitivity goals at low frequency. Moreover, the call for a second 20 km facility is motivated mainly by the necessity of measuring the post-merger oscillations of binary neutron star mergers which are expected
to happen between 2–4 kHz, where the optical response of a 40 km detector is reduced due to the
travel time of the light down the arms \cite{CE1, CE2, CE3, CE4}. In particular, the 20 km could make use of “tuned”
operation, i.e., the quantum noise is reduced in the 2–4 kHz band at the expense of higher quantum noise elsewhere.

\begin{figure}[H]
\centering
\includegraphics[width=10.0 cm]{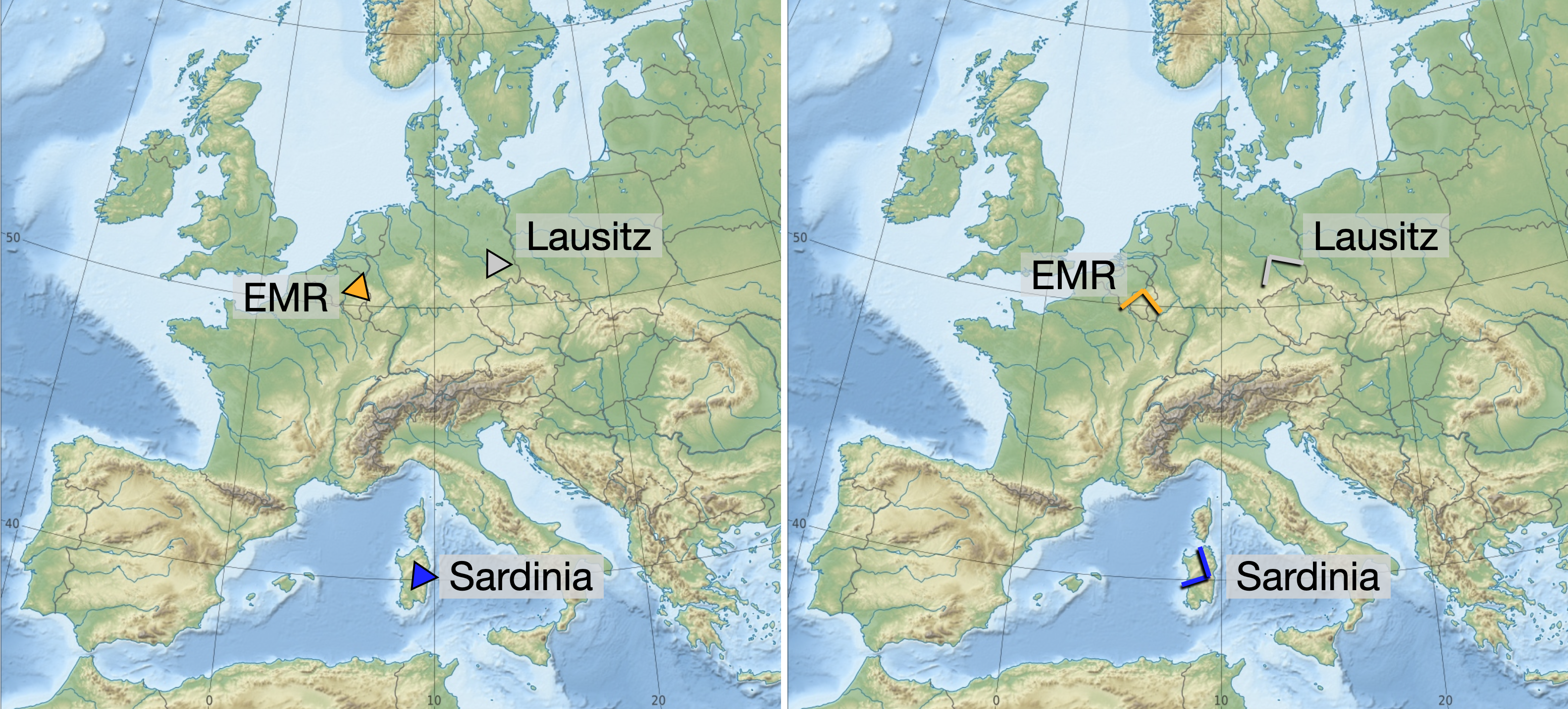}
\caption{Map of Europe showing the location of the sites candidate to host ET. (left) The triangles shown represent realistic orientations for the proposed triangular detector at each candidate site. The size is not to scale. An hypothetical triangular detector will be hosted at only one of the sites. (right) Possible orientations of the L shaped detectors. For Sardinia and Lausitz, the orientations are based on appropriate data considered by the local site characterization teams, whereas the orientation of the L for the EMR region is purely speculative. The size of the arms is not to scale. In case the 2L network will be chosen for ET, the detectors could be hosted by either the Sardinia-Lausitz or Sardinia-EMR couples, since the EMR-Lausitz baseline is too short for providing effective sky localization \cite{iacovelli26}.\label{fig:map}}
\end{figure} 

As of 2026, there are three sites which are officially candidate to host the ET triangle or one of the detectors of the 2L network (Figure \ref{fig:map}): the  Euregio Meuse-Rhine (EMR) \cite{ET2011, ET2020}, between Belgium and the Netherlands, the area surrounding the Sos Enattos former mine in Sardinia (Italy) \cite{ET2011, ET2020, naticchionietal2014, naticchionietal2020, digiovannietal2021, digiovanni2023, saccorotti23, diaferia25, Diaferia26} and the Lausitz region, in Saxony (Germany). Tables \ref{tab:et_triangle_configurations} and \ref{tab:et_l_configurations} show the proposed detector coordinates for each candidate site. Since 2010, the candidate sites sites have been the subject of thorough characterization studies
aimed at assessing their suitability to host ET \cite{naticchionietal2014, naticchionietal2020, digiovannietal2021, digiovanni2023, saccorotti23, Bader_2022, koley2022surface, diaferia25, Diaferia26} and the impact of site dependent noise on the observability of GW signals \cite{alloccaetal2021,janssens2022,janssens2024,digiovanni2025}. These studies outlined the importance of accounting for site dependent noise on the design of ET and its science goals and helped to clarify the challenges that this new and ambitious project is expected to face. 

As a consequence, the goal of this review is to give a comprehensive pedagogical overview of the challenges related to the surrounding environment that ET is expected to face, both during its design phase and its operations, and the spirit inspiring the design choices of the experiment. These challenges are paramount for the preparatory phase of ET and will drive the planning of its infrastructure. Moreover, we will not discuss the engineering details of the ET infrastructure, since they are still in their infancy, are constantly evolving and are heavily site dependent, thus making any discussion on this topic potentially obsolete in a matter of a few months. On the other hand, we to provide the essence of the background needed to support the engineering studies for the construction of ET.
\begin{table}[htbp]
\centering
\begin{tabular}{cccccc}
\hline
Site & Vertex & Latitude [deg] & Longitude [deg] & $x$-arm azimuth [deg] & $y$-arm azimuth [deg] \\
\hline
Sardinia &S1 & 40.472806 & 9.448954 & 198.984 & 138.986 \\
 &S2 & 40.441343 & 9.329487 & 79.061 & 19.062 \\
 &S3 & 40.536110 & 9.353495 & 319.048 & 259.045 \\
 \hline
EMR &T1 & 50.657500 & 5.934200 & 100.760 & 160.652 \\
 &T2 & 50.753300 & 5.905500 & 219.974 & 280.782 \\
 &T3 & 50.690210 & 5.787140 & 340.765 & 40.066 \\
  \hline
Saxony &L1 & 51.346692 & 14.297119 & 325.000 & 265.000 \\
 &L2 & 51.295078 & 14.414558 & 204.908 & 144.908 \\
 &L3 & 51.257149 & 14.284634 & 85.010 & 25.010 \\
\hline
\end{tabular}
\caption{Proposed latitude, longitude coordinates and arm azimuths for the ET triangle configurations, see Figure \ref{fig:map}}
\label{tab:et_triangle_configurations}
\end{table}
\begin{table}[htbp]
\centering
\begin{tabular}{cccccc}
\hline
Site & Points & Latitude [deg] & Longitude [deg] & $x$-arm azimuth [deg] & $y$-arm azimuth [deg] \\
\hline
Sardinia&s1$\equiv$S1 & 40.472788 & 9.448977 & 190.619 & 100.665 \\
&s2 & 40.447765 & 9.275188 & 10.732 & 55.755 \\
&s3 & 40.605530 & 9.416177 & 280.686 & 235.663 \\
\hline
Saxony &l1$\equiv$L1 & 51.346692 & 14.297119 & 325.000 & 235.000 \\
&l2 & 51.269226 & 14.473178 & 144.863 & 189.863 \\
&l3 & 51.236183 & 14.173929 & 55.096 & 10.096 \\
\hline
\end{tabular}
\caption{Same as Table~\ref{tab:et_triangle_configurations}, but for the L-shaped configurations, see Figure \ref{fig:map}. Since the official coordinates for the L detector in EMR have not been realeased, these are not shown here.}
\label{tab:et_l_configurations}
\end{table}

This paper is organized as follows: in Section \ref{sec:ground} we discuss the basics of ground motion that are needed to understand some of the principles explained in this paper; in Section \ref{sec:generic} we discuss the noise budget of ET, outlining all the noise contribution that will shape the sensitivity of ET; \ref{sec:envnoise} focuses on environmental noise contributions; in Section \ref{sec:experience} we give an overview of the current knowledge of the impact of environmental noise on the Virgo detector; in Section \ref{sec:dependent} we give an overview of how site dependent noise may have an impact on the sensitivity of ET and its observing capabilities and in Section \ref{sec:control} we summarize possible strategies under consideration to mitigate the impact of environmental noise in ET.

\section{The Einstein Telescope and its noise budget}\label{sec:generic}
A GW interferometer is an extremely complex instrument. Its operation is the consequence of an elaborate network of sub systems which have to operate properly in order to grant the full exploitation of its scientific potential. Nevertheless, each sub system has an intrinsic noise that limits the sensitivity of the detector in a unique way. As a consequence, the sensitivity of a GW interferometer in function of frequency is determined by the combined contribution of multiple noise sources that limit the ability of the instrument to measure differential changes in the path length of the light traveling in its arms. This combined contribution is often referred to as the noise budget.

\begin{figure}[H]
\centering
\subfloat[\centering]{\includegraphics[width=7.0cm]{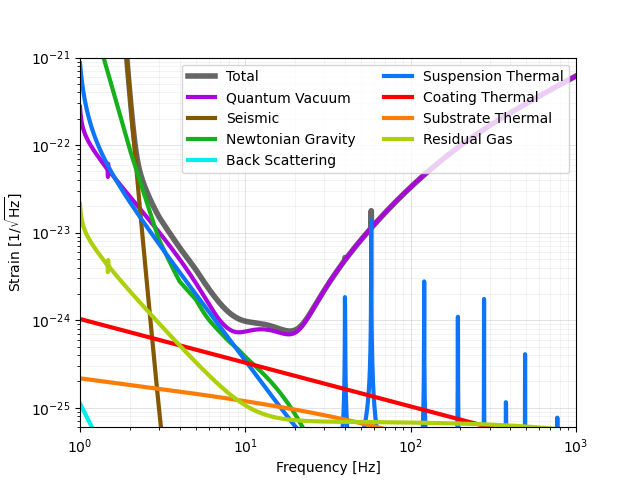}}
\subfloat[\centering]{\includegraphics[width=7.0cm]{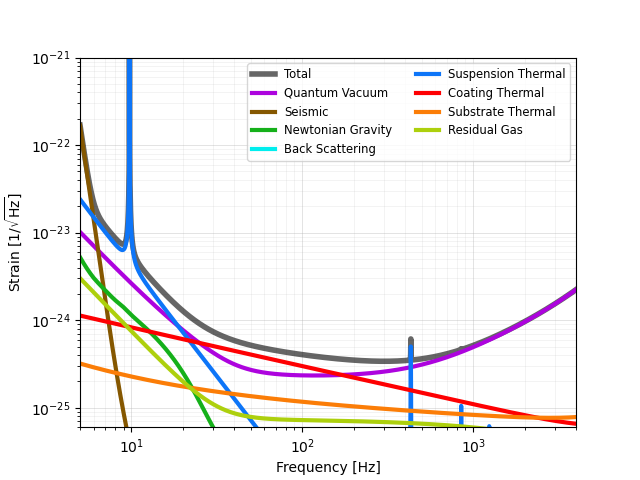}}\\
\caption{Example of a Noise budget of the ET LF (a) and HF (b) detectors for the ET triangular configuration. The noise budget is calculated using pyGWINC \cite{pygwinc_ascl}. \label{fig:NB}}
\end{figure}

Since ET foresees the employment of a so-called xylophone configuration, in which two interferometers optimized for different frequency bands operate in parallel, its noise budget will differ between the LF and HF interferometer (Figure \ref{fig:NB}) and the overall sensitivity of the detector is given by the combined sensitivity of the two detectors:
\begin{equation}
    \label{eq:NB}
    S_{\mathrm{ET}} = \frac{1}{\frac{1}{S_{\mathrm{LF}}}+\frac{1}{S_{\mathrm{HF}}}}
\end{equation}

\begin{figure}[H]
\centering
\subfloat[\centering]{\includegraphics[width=7.0cm]{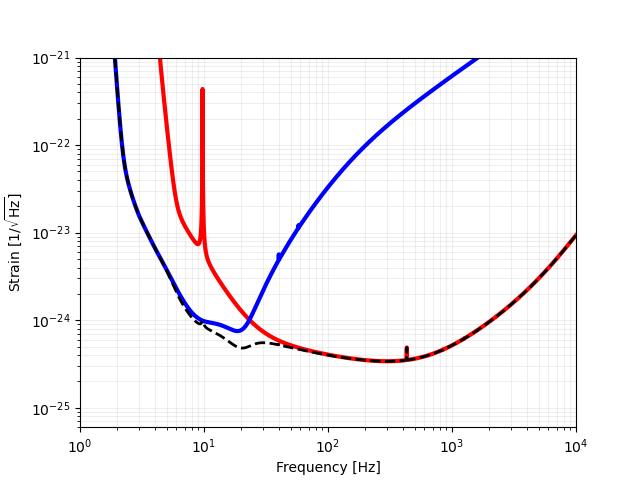}}
\subfloat[\centering]{\includegraphics[width=7.0cm]{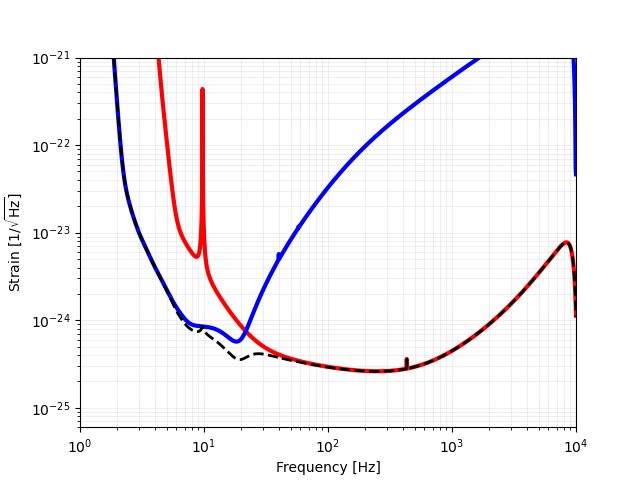}}
\caption{The sensitivity curve of ET (black) is a composition of the LF (blue) and HF (red) sensitivity curves combined according to Equation \ref{eq:NB}. (a) stands for the triangular configuration. (b) is an example of the L  shaped detector with 15 km long arms. The sensitivity curves reported here are from the same models as Figure \ref{fig:LVKETsens}. For more detailed comparisons between the potentials of triangle and 2L detectors, please refer to \cite{coba, Iacovelli_2024}. \label{fig:deltanb}}
\end{figure} 

Noise sources of a GW interferometer can be divided into classes according to their origins and coupling mechanisms. A first distinction consists into splitting them into displacement and sensing noises: displacement noises are those that cause a real motion of the test masses or their surfaces; sensing noises limit the instrument's ability to measure test mass motion. Nevertheless, this distinction is not perfect, since some noise sources can belong to both categories. In addition to that, it is common practice to classify the noise also as fundamental, technical, and environmental. Fundamental noises can be computed from first principles, and they determine the ultimate design sensitivity of the instrument. It is not possible to reduce this kind of noise without a major instrument upgrade, such as the installation of a new laser or the fabrication of better optical coatings for the mirrors. Technical noises, on the other hand, arise from electronics, control loops, charging noise and other effects that can be reduced once identified and carefully studied. Given the leap in sensitivity represented by ET, the mitigation and control of technical and environmental noises is of the uttermost importance for the appropriate functioning of the detector. Environmental noises, which represent the main topic of this paper, include seismic motion, acoustic and magnetic noises. Environmental noise sources will be extensively discussed in Sections \ref{sec:ground} and \ref{sec:envnoise}. In this section we give an overview of the remaining noise contributions that shape the characteristic sensitivity curves of GW interferometers (Figure \ref{fig:NB}).
  \begin{itemize}
    \item \textbf{Quantum noise} - affects the entire bandwidth of GW interferometers. It is the noise which originates from the quantum nature of light \cite{saulson94, squeezing3}. 
    Notably, quantum noise is the balance of two sources of error: the uncertainty in the number of photons, whose distribution is ruled by a Poisson statistic, that are collected by the photodiode at the output of the detector (\textit{shot noise}); a fluctuating force exerted on the mirrors due to the photon number fluctuations that pushes the test masses in random a manner that can mask their GW induced motion, (\textit{pressure radiation noise}) \cite{radpress}. As the input laser power increases, the photon-counting error decreases, whilst increasing the radiation-pressure error at low frequency. The photon counting error can be reduced by the adoption of quantum squeezing \cite{squeezed, squeezing1, squeezing2, squeezing3}. \textbf{The goal of this technology is to reduce quantum uncertainty in one observable (e.g., position) at the expense of increased uncertainty in its conjugate (momentum), allowing detectors to selectively suppress either shot noise or pressure radiation noise}. On the other hand, increasing the mass of the suspended mirrors lowers their response under fluctuating radiation pressure forces.
    \item \textbf{Thermal and Brownian noises} \cite{Puppo2011} - These noise sources originate from mechanical systems (mirrors, suspensions, glass fibers, coating of the mirrors) that can be seen as dissipative oscillators and subject to random thermal fluctuations. The thermal equilibrium with the external environment will make them vibrate, causing an uncertainty in the measurement of their position according to the fluctuation-dissipation theorem \cite{teoF&D}. This noise can be reduced using mechanical systems with low internal dissipations or, as in the case of KAGRA, using mirrors at cryogenic temperatures \cite{kagracryo1, kagracryo2}.
    \item \textbf{Residual gas} \cite{sinica} - Refers to the noise generated by residual gas in the vacuum system. This gas interacts with the main laser beam, scattering light into unwanted directions. The scattered photons can then bounce off vibrating surfaces and re-enter the main beam, creating a noise signal that can either mimic a GW or interfere with the detector's control system. For this reason, all critical components are housed in ultra-high-vacuum enclosures to eliminate noise from residual gas motion and acoustic coupling as well;
    \item \textbf{Back scattering:} \cite{Longo2026} - Refers to the photons that, scattered by surfaces in the optical path, couple with the main laser beam.
    \item \textbf{Seismic noise}: - Refers to the residual ground motion not filtered by the suspension isolation systems and transferred to the test masses of the detector. It causes uncertainty in the measurement of the path length differential variation. It will be discussed in Section \ref{sec:envnoise}.
    \item \textbf{Newtonian Noise}: - Also known as gravity gradient noise. It is caused by local mass density fluctuations which alter the gravitational potential in the vicinity of the test masses. It will be discussed in Section \ref{sec:envnoise}.
  \end{itemize}

\section{An overview of ground motion}\label{sec:ground}
Before proceeding any further and discussing the effect of environmental noise on a GW detector, it is appropriate to summarize some of the key concepts of ground motion that are at the basis of the noise sources discussed in Section \ref{sec:envnoise}. The performance and operations of ground-based GW detectors are affected by ground motion which generates seismic noise. This noise affects the sensitivity in the range 2 Hz - 10 Hz, often threatening the duty cycle of the detector. In the following, we will focus on seismic noise in that frequency range, however, it is worth pointing out that significant excess noise below this range, albeit out of the ET accessible frequency band, can also impact the interferometer controls and have negative effects on the sensitivity of the detector. To be more specific, low-frequency seismic disturbances can couple directly into the suspension systems and active isolation controls. This coupling generates control forces that inject electronic noise or saturate the actuators, thereby degrading the duty cycle of the detector and the overall sensitivity. As a consequence, ground motion for GW detectors has to be monitored all the way from the lowest bound of seismic noise $(\mathcal{O}(\mathrm{mHz}))$ up to 10 Hz. On top of that, in the very-low frequency range, standard inertial sensors cannot distinguish between true horizontal translational acceleration and ground tilt, a fundamental limitation known as the tilt-translation coupling phenomenon \cite{difronzo2026}. Consequently, a significant increase of horizontal ground motion at $(\mathcal{O}(10-100\,\mathrm{mHz}))$ could be interpreted as an excess of environmental tilt noise; the interferometer control systems could erroneously move the test masses to compensate for a tilt effect and generate an excess of control noise in the low frequency band of the detector. As a consequence, GW detectors need sensors capable of measuring the inclination of the ground surface and specifically designed to minimize the coupling with ground translations, in order to be mainly sensitive to ground tilts \cite{tilt1, tilt2, tilt3, tilt4}. For a comprehensive overview on how to address tilt in GW detectors, see Ref. \cite{difronzo2026}.

Since the Earth is a live planet, with continuous human and natural activities, the ground is permeated by a persistent and variable vibration that produces a spectrum with a characteristic shape, with some variations according to the peculiar features of the site where it is measured. These vibrations are generated by seismic waves traveling through the Earth’s ground layers at different speeds and produced by the most varied phenomena, e.g., wind, ocean waves, earthquakes, human activities. Seismic wave speeds can change depending on the density and the elasticity of the crossed medium, as well as on their depth. Seismic waves are classified in the following way:\cite{potsdam}
\begin{itemize}
    \item \textbf{Body waves:} they propagate through media in all directions. The propagation velocity depends on the density and elasticity of the medium that they cross. This dependence is due to lithostatic pressure which increases with depth, thus compressing rocks and closing microcracks, significantly enhancing rock stiffness and elasticity and increasing propagation velocity. Velocity jumps can occur at  discontinuities, where rock properties change.
    \item \textbf{Surface waves:} as the name implies, they travel only on the Earth's surface and their amplitude exponentially decays with depth $z$ \cite{potsdam} as $A(z) = A_0 e^{-\frac{z\omega}{v_{\mathrm{surf}}}}$, where $v_{\mathrm{surf}}$ is the speed of surface waves. During an earthquake, the displacement amplitude of surface waves can be up to several cm. Surface waves are generated by the complex interaction of body waves with the discontinuity represented by the Earth's surface.
\end{itemize}

The nature of body waves is described by the equation of ground motion. If we consider a point outside the source region, in homogeneous media and for small deformations, the equation of motion is\cite{potsdam}
\begin{equation}\label{eq:gmotion}
    \rho \ddot{\vec \xi} = (\lambda + 2\mu) \nabla \nabla \cdot \vec \xi - \mu \nabla \times \nabla \vec \xi,
\end{equation}
where $\vec\xi$ is the displacement vector, $\rho$ is the ground density, $\lambda$ and $\mu$ are two parameters known as Lamé parameters. Specifically, $\mu$ represents the shear modulus. It quantifies a material's resistance to shear deformation and changes in shape. On the other hand, $\lambda$ is related to volumetric compression and expansion through the bulk modulus defined as $K \equiv \lambda + \frac{2}{3}\mu$. From the mathematical nature of the terms of Equation \ref{eq:gmotion}, we can infer that the divergence term implies a compression/decompression, whereas the curl term represents a change of shape without change of volume (pure shearing).\cite{potsdam} In fact Equation \ref{eq:gmotion} has two independent solutions that represent body waves, which are compressional waves (P waves) and shear waves (S waves) waves respectively:\cite{potsdam}
\begin{equation}
    \frac{\partial^2 (\nabla\cdot \vec \xi)}{\partial t^2} = \frac{\lambda + 2 \mu}{\rho}\nabla^2(\nabla\cdot\vec \xi),
\end{equation}
\begin{equation}
    \frac{\partial^2 (\nabla \times \xi)}{\partial t^2} = \frac{\mu}{\rho}\nabla^2(\nabla\times\vec \xi).
\end{equation}
The propagation velocities are\cite{potsdam}
\begin{equation}
    v_\mathrm{P} = \sqrt{\frac{\lambda+2\mu}{\rho}} = \sqrt{\frac{K+\frac{4}{3}\mu}{\rho}},
\end{equation}
\begin{equation}
    v_\mathrm{S}\sqrt{\frac{\mu}{\rho}},
\end{equation}
with $v_\mathrm{P}>v_\mathrm{S}$. As shear and bulk moduli and $\rho$ change with depth, body wave velocity increases and discontinuities in subsoil properties lead to velocity jumps. As already mentioned, S waves lead to a change of shape without a change in volume.\cite{potsdam} This implies that S waves do not cause a change in local ground density, unless they meet a cavity (e.g., a cavern), in which case they alter the local density due to the vibrations induced on the walls of the cavity.

On the other hand, in the presence of a free surface, other solutions are possible and describe surface waves. Surface waves are divided in Love waves and Rayleigh waves. The former are generated by the constructive interference of repeated reflections at the free surface of horizontal S waves generated by a earthquakes that are very far away from the observer (teleseisms).\cite{potsdam} They can also be generated by the constructive interference between horizontal S waves which are reflected within a homogeneous layer overlaying a half-space with higher propagation velocity.\cite{potsdam} Rayleigh waves originate from the inhomogeneous coupling of vertical S waves and P waves at the free space and are slower than Love waves.\cite{potsdam}

\begin{figure}[H]
\centering

\subfloat[\centering]{\includegraphics[width=7.0cm]{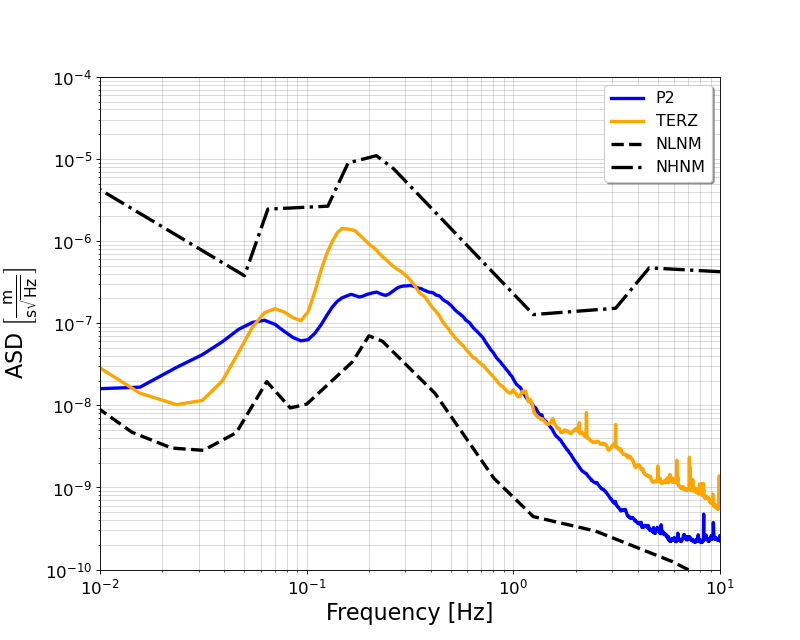}}

\subfloat[\centering]{\includegraphics[width=7.0cm]{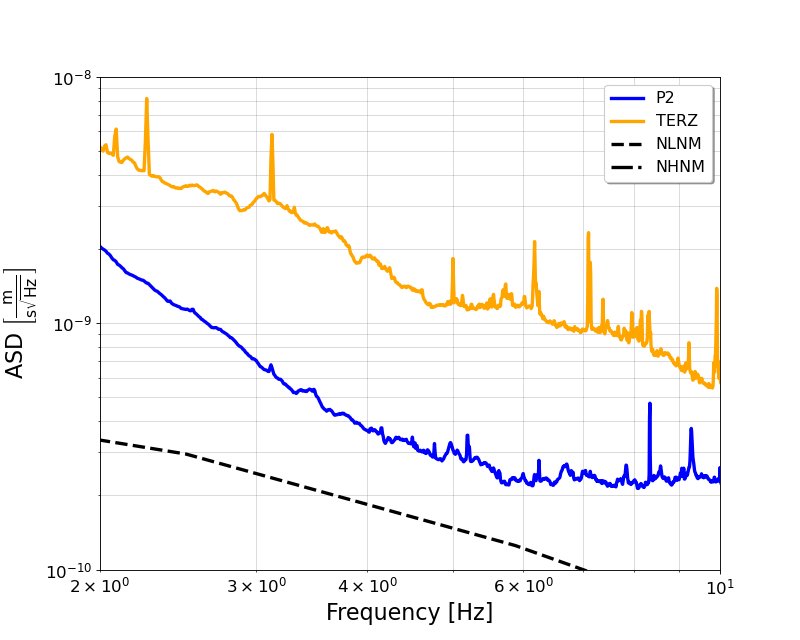}}\\

\caption{(a) Average seismic spectra obtained from the Sardinia (blue) and EMR (orange) candidate sites at the target depth for ET ($\simeq 250\,\rm m$). The NLNM and NHNM are shown. Some of the seismic spectra features are clearly visible, such as the primary and secondary microseismic peaks, albeit with some differences in shape, peak frequency and prominence, due to local effects. The quietness of the Sardinia site in the frequency band of interest for ET (f $>$ 2 Hz) is clearly testified by the spectra almost touching the NLNM. Spectra below 0.01 Hz are not shown, since the seismometers are not calibrated at those frequencies. (b) Zoom of the seismic spectra in the 2 Hz - 10 Hz range, relevant for ET. The difference in seismic levels is up to one order of magnitude, to the advantage of Sardinia. This has an impact over the potential performance of ET, as outlined in Section \ref{sec:dependent}. \label{fig:spectracomp}}
\end{figure}

As mentioned above, the persistent vibrations of the Earth recorded by a seismometer generate a spectrum with a characteristic shape (Figure \ref{fig:spectracomp}). To each frequency band, a particular noise source can be associated.
\begin{itemize}
    \item \textbf{Seismic hum }$[\mathbf{f \lesssim 20\,\mathrm{mHz}}]$: can be observed anywhere on Earth and originates from the beating of the oceans on the sea floor;
    \item \textbf{Tilt noise }$[\mathbf{50 \, \mathrm{mHz}\lesssim f \lesssim  500\,\mathrm{mHz}}]$: site dependent as well as the seismic hum, according to the local properties of the soil, and is generated by the ground tilting due to atmospheric pressure;
    \item \textbf{Microseismic noise }$[\mathbf{500 \, \mathrm{mHz}\lesssim f \lesssim  0.5\,\mathrm{Hz}}]$: observable everywhere on Earth, even though the spectral properties depend from the observer's site. This noise is generated by sea waves as well as the seismic hum. However, the mechanism is different. In particular, at $f\simeq700\,\mathrm{mHz}$ we observe the primary microseismic peak, which is associated with the interaction of waves and scaling sea floor near the shores. At $f\simeq0.2\,\mathrm{Hz}$ we observe the secondary microseismic peak, which originates by the beating on the sea floor of a standing wave field originated by sea wave trains traveling along opposite directions. The shape and prominence of these peaks strongly depends on the distance of the observer from the sea, the seasonality and on other local features, as seen in Figure \ref{fig:spectracomp}.
    \item \textbf{Microseismic noise tail }$[\mathbf{0.5 \, \mathrm{Hz}\lesssim f \lesssim  5\,\mathrm{Hz}}]$: in this frequency band, seismic noise amplitude slowly decays and is mainly affected by natural sources such as wind or by catastrophic atmospheric events.
        \item \textbf{Anthropic noise }$[\mathbf{f > 5 \mathrm{Hz}}]$: in this frequency band, seismic noise from anthropic sources and activities dominates and is mostly site dependent.
\end{itemize}
\subsection{Low and High seismic noise models}
Given the varying nature of seismic spectral features according to the site and with the purpose of monitoring these variations, in the early 1990s, J. Peterson \cite{peterson1993observations} collected and analyzed data from seismic stations located all around the world. From the data he inferred the higher and lower envelopes of the spectra and derived the New High Noise Model (NHNM) and the New Low Noise Model (NLNM). These models represent a global statistical property of seismic noise, according to which the NLNM and NHNM are the 10th and 90th percentile, respectively, of the global distribution of seismic spectra. These models are commonly used as a reference for the lowest and the highest seismic background spectra that it is possible to find on the Earth and. As a consequence, a site is considered quiet from a seismic point of view if its seismic spectra are close to the NLNM model. With this in mind, Figure \ref{fig:spectracomp} outlines the quietness of the Sardinia site compared to the EMR. This also suggests that, given the difference in noise levels, environmental noise mitigation strategies, which will be discussed later in the paper, will likely be site dependent and should be designed according to the peculiar features of environmental noise recorded at each candidate site.

\section{Environmental noises}\label{sec:envnoise}
As already mentioned, with respect to current detectors, ET is aiming for the extension of the accessible frequency band down to 2\,Hz thanks to the presence of the cryogenic LF interferometer. The extension of the observational bandwidth and the huge leap in sensitivity in the LF range comes at the cost of being more prone to several types of noise sources caused by the surrounding environment. Some of these noise sources directly affect the sensitivity and duty cycle of the machine, whereas others act on the control systems of the apparatus. Generally speaking, environmental disturbances couple with the detector output through both linear and non-linear mechanisms (Table \ref{tab:noise_sources}) \cite{cvse,Fiori2020,o3noise}. In several cases, weak environmental excitations can produce significant excess noise due to up-conversion mechanisms or scattered light processes. As a consequence, the identification of the physical coupling paths is often considerably more complex than the simple identification of a spectral line in the strain channel. Dedicated noise hunting activities are needed to correlate environmental data with the detector output and to identify the origin of noise transients, spectral lines and other non-stationary noises. 

\begin{table}[htbp]
\centering
\begin{tabular}{ccccc}
\hline
Noise source & Field type & Coupling & Affected subsystem & Effect \\
\hline
Train                      & seismic          & ground $\rightarrow$ suspension  & test masses        & displacement    \\
Train                      & magnetic         & EM field $\rightarrow$ magnets   & controls         & glitches    \\
Wind                       & seismic/acoustic & ground/air           & suspension/optics  & excess noise    \\
Solar storm                & magnetic         & field $\rightarrow$ actuators    & controls            & glitches        \\
Earthquake                 & seismic          & ground $\rightarrow$ suspensions & interferometer     & loss of lock    \\
Mass density\\ fluctuation & gravity    & direct               & test masses          & NN              \\
\hline
\end{tabular}
\caption{Examples of possible external environmental noise sources and coupling mechanisms. Noise sources related to the detector infrastructure are not reported here for simplicity}
\label{tab:noise_sources}
\end{table}

In this section we give a summary of the four main classes of environmental noises which are expected to play a crucial role in the operations of ET, i.e., seismic, acoustic, magnetic and Newtonian noise, whereas a short review of coupling mechanisms and other transient noise sources related to the infrastructure of a GW detector is given is Section \ref{sec:experience}. In this regard, in the selection process for the sites to host ET, it is vital to carefully assess local natural and anthropogenic noise sources that can impact the mitigation strategy adopted to reduce noise in the detector, as well as the cost of the development of the infrastructure.\cite{amann} This is particularly relevant in the case of the ET triangular configuration, where locally coincident environmental disturbances may reduce the effectiveness of coincident signal detection, especially for weak GW events.

\subsection{Seismic Noise}
Seismic noise is a displacement noise that acts directly on the interferometer in the low frequency range. Through ground motion (Figure \ref{fig:spectracomp}), it induces unwanted vibrations and motions of the test masses and other optical components. This motion does not only generate uncertainty in the measurement of the length of the optical path but can also deviate the laser light from its intended path and induce light scattering  from seismically excited surfaces which couples back into the main beam of the interferometer \cite{accadia2010noise}. 
Sources of seismic noise are both natural and human-made. If the former can span a wide range of frequencies, from a few mHz up to 100 Hz, the latter appear above a few Hz \cite{bonnefoy2006nature}. Natural noise sources can originate from, e.g., earthquakes, sea activity and local winds, whereas human-made sources include all human activities outside (e.g. roads, trains and industrial activities, infrastructures-ground interaction, wind farms) and inside the laboratory (e.g. self-inflicted vibrations due to fans and air conditioning), as it will be discussed in Section \ref{sec:experience}.. Microseismic activity (in the range of about 0.1–-1\,Hz) is caused by sea wave perturbations acting on the seafloor and on the shore. Depending on the distance from the sea/ocean and the ground geology, microseismic peaks can be larger or narrower. To better control the effects of microseismic noise on a GW interferometer, the KAGRA collaboration has developed a method to foresee the microseismic noise levels based of weather forecast in the vicinty of the detector itself \cite{hoshino}. The transition between seismic and anthropic contributions occurs between 0.5 and 5\,Hz \cite{digiovanni2023}. Wind can interact with surface obstacles (such as buildings and trees), creating ground vibrations \cite{digiovanni2023} and tilts that can be detected by an interferometer. Other local environmental sources, such as bridges, human structures and wind farms, can be identified as peaks in the seismic spectrum \cite{saccorotti, digiovanni2023, saccorotti23, diaferia25}. Special attention is given to wind farms, as it was found that the seismic signature of these facilities several km away from the LIGO and Virgo detectors is still clearly visible \cite{ligowind, saccorotti}. The case of wind farms will be discussed in Sections \ref{sec:experience} and \ref{sec:control}.

To tackle ground motion properly, appropriate suspension systems for the test masses are needed. Current detectors such as LIGO, Virgo and KAGRA, followed different approaches. LIGO adopted an active isolation platform combined with a quadruple pendulum to which the test masses are suspended.\cite{ligosusp} On the other hand, Virgo \cite{Ballardin2001VirgoSuperattenuator, superatt} and KAGRA \cite{kagrasusp} chose a purely passive isolation system which consists of a 8 stage pendulum, in the case of Virgo, attached to an inverted pendulum \cite{invertedP} that acts as a pre-isolator stage which also allows for precise control of the mirrors. 

\begin{figure}[t]
\centering
\includegraphics[width=7.0 cm]{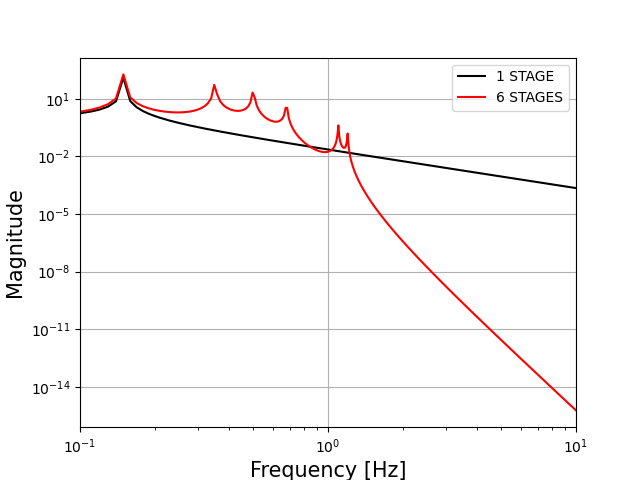}
\caption{Comparison of the transfer function for a single pendulum (black curve) and a 6 stage pendulum (red curve) using Equation \ref{eq:TF2} and the Y-axis represents a dimensionless magnitude. The benefits of using  a multi stage pendulum for sesimic isolation in the frequency range of GW detectors is apparent.\label{fig:TF}}
\end{figure}

To better understand the use of multi stage pendulums for seismic isolation, let us consider the modulus of the transfer function for a simple pendulum subject to friction is
\begin{equation}\label{eq:pend}
    |H(\omega)| = \frac{\sqrt{1+\phi}}{\sqrt{\left (1-\frac{\omega^2} {\omega_p^2}\right)^2+\phi^2}},
\end{equation}
with $\phi$ being the dissipation factor of the system and $\omega_p = 2\pi f_p$ the resonant frequency. 
From Equation \ref{eq:pend} in is clear that a pendulum acts as a low pass mechanical filter for which at $f<<f_p$ there is total transmission, amplification for $f=f_p$ and attenuation for $f>>f_p$. If we consider a N stage multiple pendulum,
\begin{equation}\label{eq:TF2}
    |H(\omega)| = \prod_{i=1}^{N} \frac{\sqrt{1+\phi_i}}{\sqrt{\left (1-\frac{\omega^2}{\omega_i^2}+\phi_i^2\right)}}.
\end{equation}
 This implies that, for $f>>f_1$ the attenuation is even more effective. As a consequence, given a multiple pendulum, with each of the N stages having a resonant frequency $f_1>f_2>\ldots>f_N$, frequencies $f>f_1$ will be attenuated by a factor
\begin{equation}
    \mathrm{A} = \frac{1}{f^{2N}}
\end{equation}
and make this device an excellent isolation system for a GW interferometer (Figure \ref{fig:TF}). The adoption of the inverted pendulum pre-isolator stage contributes in making the attenuation even more effective. The excellent performance of the passive isolation system in Virgo pushed the GW community to consider an evolution of system for ET, able to effectively mitigate underground ground motion down to 2 Hz.

\subsection{Acoustic and Infrasound Noise}\label{sec:acoustic}
Acoustic noise, air pressure perturbations and vibrations can be produced by local sources such as wind, trains and fans \cite{infra}. Acoustic noise is relevant as long as it couples with the subsystems of the detector. For example, fans with a rotation frequency between 20 Hz and 30 Hz can produce acoustic induced vibrations that affect external optical benches. On other occasions, air vibrations can couple with the ground generating an excess of seismic noise or, conversely, ground motion can induce air vibrations that propagate in the ambient and couple with the detector sub systems in the experimental area. Since ET is foreseen to be built underground and will be provided with all the devices needed to create an habitable environment (e.g., air-condition and air circulation systems), acoustic noise is regarded as one of the major sources of noise that can affect the sensitivity of the detector. Its mitigation strategies are part of the engineering effort to build a scientifically sustainable infrastructure underground.

Another potentially significant noise source is atmospheric infrasound. Infrasound waves are defined as acoustic waves
below 20 Hz and constitute a persistent component of the ambient atmospheric noise background \cite{mariusz}. Infrasound waves produce fluctuations in air density that generate time-varying gravitational forces acting on the detector test masses through NN. Moreover, infrasound waves can couple with the ground to generate an excess of seismic noise as in the case of wind turbines \cite{saccorotti} (Sections \ref{sec:experience} and \ref{sec:control}). The magnitude of infrasound depends on various site dependent elements, including atmospheric and weather conditions, topography and underground depth. Underground environments offer significant advantages, as they provide natural shielding from surface disturbances and reduce the coupling of atmospheric and seismic noise to the detector.

\subsection{Magnetic Noise}\label{sec:magnetic}
\begin{figure}[H]
\centering
\subfloat[\centering]{\includegraphics[width=8.0cm]{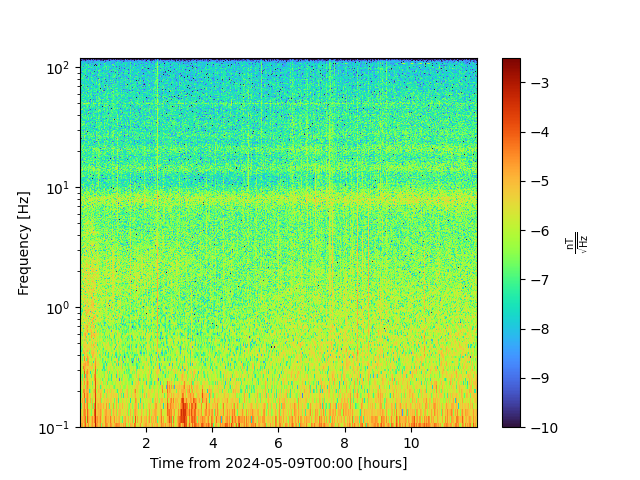}}
\subfloat[\centering]{\includegraphics[width=8.0cm]{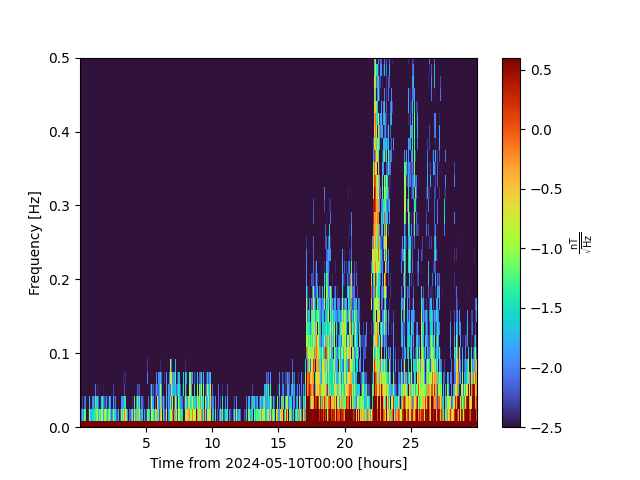}}\\
\caption{Examples of time frequency plots of magnetic noise of natural origin from the Sardinia candidate sites. Both plots are obtained from data from a sensor placed at a depth of 111m in the Sos Enattos mine. (a) Sardinia is generally very quiet from a magnetic point of view and Schumann resonances are clearly visible. (b) Example of a noise transient of natural origin which affects the low frequency part of the magnetic spectrum. This type of noise is related to solar storms which are also visible in sensors placed all around the globe.\label{fig:magnoise}}
\end{figure}

Another important noise contribution which is expected to have an impact over the ET-LF sensitivity and control systems is the magnetic noise. In fact, a GW interferometer is a complex machine which, to be operated properly, is provided with magnetic components (e.g., magnetic actuators of the mirrors, magnetic anti-springs in the stages of the cascade pendulum to mitigate the vertical ground motion) which are susceptible to unwanted magnetic fields. These fields can be of natural ,e.g. Schumann resonances, which are caused by electromagnetic signals excited in the cavity formed between the Earth’s surface and the ionosphere (Figure \ref{fig:magnoise}.a), or human-made (any device which carries al electric current) origin. The latter can be distinguished between those which enter the interferometer from outside (e.g. those produced by trains or power lines) and those produced by the interferometer itself (e.g. electronics or the power grid). All lessons learned with the LIGO and Virgo experiments, briefly discussed in Section \ref{sec:experience}, will be taken into account in the construction of ET.  However, studies of magnetic noise coupling functions for the ET detector, measurements of underground magnetic noise at the candidate sites and the mitigation strategies are all still work in progress.
For what concerns magnetic noise of natural origin, it is worth pointing out that the Sun is another potential source of magnetic noise. Solar activity disturbs our magnetosphere and the effects are visible both on the surface of the Earth and underground (the effects of a solar storm detected by and underground magnetometer are visible in Figure \ref{fig:magnoise}.b). Events with a strong impact on the Earth's surface occur only a few times per year during solar maximum (solar activity has an 11-year cycle, or 22 years when considering the change of polarity). On the other hand, Magnetic perturbations impact different geomagnetic locations in different ways: noise levels are higher at higher geomagnetic latitude and, since the frequencies are below a few Hz, they can penetrate almost unperturbed underground. Solar events can impact the measurements directly (observable up to a few Hz) and the control system of the detector.

As a consequence, magnetic noise is another key aspect which is taken into account during the site selection process for ET. In fact, the first necessary step is selecting an intrinsically electromagnetically quiet site, while implementing effective preservation strategies to maintain its low-noise conditions over time. The goal of minimizing the site-infrastructure induced noise requires careful planning of mitigation measures already during the site design phase.

\subsection{Newtonian Noise}\label{sec:newtonian}

Direct coupling of Seismic Noise with test masses of a GW interferometer through ground motion is strongly attenuated by the suspension systems that reduce the residual motion to acceptable levels. In the ideal case, we could build the perfect suspension that filters out all of ground motion to keep the test masses perfectly still. Nevertheless, ground motion generates another type of noise, known as Newtonian noise,\cite{harms} caused by the local soil density fluctuations induced by ground motion which change the gravitational potential experienced by the test masses. This causes a motion of the test masses due to the fact that the direction of $\vec g$ changes with time. A similar effect is generated by acoustic noise producing density fluctuations in the atmosphere or inside the detector infrastructure. The effects of Newtonian noise over the operations of GW detectors in the low frequency range are well known since the first assessment of the noise budget of an interferometer. \cite{Weiss:1972, beccaria1998relevance, Hughes1998SeismicGravityGradient} Newtonian noise is particularly tricky to be addressed, since it cannot be either mechanically shielded or attenuated and its effect, once effectively measured, can only be subtracted from the detector output \cite{cella,harms15,Coughlin_2016, badaracco2019, Badaracco2020, Koley2024}. As a consequence, Newtonian noise cancellation requires a perfect knowledge of the noise field around the test masses. Current detectors are not yet affected by Newtonian noise, however, the sharp increase in sensitivity of ET will make this noise source of key importance to reach the sensitivity goals at low frequency.

Generally speaking, since the local matter density at $\vec r_0$ depends on time $\rho (\vec r_0, t)$, these fluctuations are a direct consequence of the Newtonian potential

\begin{equation}\label{eq:potential}
    \delta \phi (\vec r_0, t) = -G \int_V \frac{\rho (\vec r, t)}{|\vec r_0 - \vec r|}dV.
\end{equation}

Moreover, since the density fluctuations are caused by ground motion ($\delta\rho_{\mathrm{seis}}(\vec r, t)$), atmospheric density variations ($\delta\rho_{\mathrm{press}}(\vec r, t)$) \cite{Creighton_2008} as well as thermal fluctuations ($\delta\rho_{\mathrm{therm}}(\vec r, t)$), $\rho (\vec r, t)$ is the combination of more terms:

\begin{equation}\label{eq:rhoseis}
    \delta\rho_{\mathrm{seis}}(\vec r, t) = -\nabla\cdot [\rho_{\mathrm{soil}}(\vec r)\xi(\vec r, t)];
\end{equation}
\begin{equation}
    \delta\rho_{\mathrm{press}}(\vec r, t) = \frac{\bar\rho_{\mathrm{atm}}}{\gamma\bar p_{\mathrm{atm}}}\delta p_{\mathrm{atm}}(\vec r, t);
\end{equation}
\begin{equation}
    \delta\rho_{\mathrm{therm}}(\vec r, t) = \frac{\bar \rho_{\mathrm{atm}}}{\bar T_{\mathrm{atm}}}\delta T_{\mathrm{atm}}(\vec r, t).
\end{equation}
In these equations, $\bar\rho_{\mathrm{soil}}$, $\bar p_{\mathrm{atm}}$ and $\bar T_{\mathrm{atm}}$ represent the average atmospheric density, pressure and temperature, respectively; $\gamma$ is the adiabatic index; $\xi(\vec r, t)$ represents ground motion. 

The density fluctuations equations make clear that the nature of Newtonian noise is closely tied to the geological, atmospheric and hydrological properties of the site of ET. As a consequence, these properties must be considered as an integral part of the experiment itself and not as a mere supplementary information needed for engineering studies. Underground geology affects the way in which the site responds to ground motion and therefore the amount of Newtonian noise related to ground displacement. On the other hand, the hydrological environment is related to the underground motion of water masses, due to e.g. rainfall, groundwater, underground water flow, which can gravitationally interact with the test masses of ET. Similar is the effect due to atmospheric pressure and temperature fluctuations. However, having a GW detector in an underground environment has some advantages. As we have already seen, surface seismic waves decay exponentially with depth and marginally contribute to the overall budget of NN. On top of that, recent investigations about the contribution of underground atmospheric pressure variations revealed that they are not expected to have a significant impact on the overall NN budget \cite{mariusz}. As a consequence and for the sake simplicity, in this work we will discuss only the density fluctuations caused by ground motion from body waves.

Inserting Equation \ref{eq:rhoseis} in \ref{eq:potential}, we get the gravity contribution to the perturbation of the Newtonian potential

\begin{equation}\label{eq:anotherpotential}
    \delta \phi (r_0,t) = G \int_V \rho(\vec r)\xi(\vec r,t)\cdot\frac{\vec r-\vec r_0}{|\vec r - \vec r_0|},
\end{equation}
from which we infer the corresponding perturbation of gravity acceleration
\begin{equation}\label{eq:acc}
    \delta \vec a (r_0,t) = -G\int_V \frac{\rho(\vec r)}{|\vec r - \vec r_0|^3} \left [\xi(\vec r,t)-3 \left ( \frac{\vec r - \vec r_0}{|\vec r - \vec r_0|}\cdot \xi(\vec r,t) \right )\frac{\vec r - \vec r_0}{|\vec r - \vec r_0|}\right ] dV.
\end{equation}
If we consider a plane seismic wave propagating in a homogeneous medium, in presence of a spherical cavern of radius R; the contribution of P waves through bulk and surface displacement; the contribution of S waves through surface contribution only (Section \ref{sec:ground}), combining Equations \ref{eq:gmotion} and \ref{eq:acc}, the acceleration perturbation at the center of the cavity is \cite{harms}
\begin{equation}\label{eq:acc2}
    \delta \vec a(\vec 0,t) = 4\pi G \rho_0 \left ( 2\xi^P(\vec 0,t)\frac{j_1(k^PR)}{k^PR}- \xi^S(\vec 0,t)\frac{j_1(k^SR)}{k^SR} \right ),
\end{equation}
with $j_1$ being the Bessel function of the first order and $k^{P/S}$ is the absolute value of the wave vector. Please, note that P waves contribute by a factor 2 higher than S waves. This highlights the fact that P waves contribute to the acceleration perturbation through both compression and decompression of the medium and through displacement of the cavern walls. On the other hand, S waves contribute only through displacement of the seismic walls. Equation \ref{eq:acc2} also outlines the fact that the description of gravity perturbation cannot overlook the composition of the seismic field in term of P and S waves.

From Equation \ref{eq:acc2} we can derive a simple analytic expression to evaluate the NN contribution to the noise budget of a GW interferometer.
First, we consider the strain of a GW interferometer as the fractional difference in path length of the arms of the detector
\begin{equation}\label{eq:A1}
    h = \frac{\Delta L}{L},
\end{equation}
Where $\Delta L$ is the difference in length of the arm $2(x_1-x_2)$ seen by a light beam traveling back and forth. At this point, for the estimation of the NN contribution to the GW strain, we consider only the $\Delta L$ due to the ground motion field $\xi (\vec r, t)$, which we assume to be stationary, homogeneous and not correlated between test masses. As such, the power spectral density (PSD) can be summarized as
\begin{equation}
    S_{h} = S_{\frac{\Delta L}{L}} = \frac{4}{L^2} S_{x_1-x_2}
\end{equation}
Now, we take account of the acceleration produced on a test mass at the center of a spherical cavern in an infinite homogeneous medium by ground motion and rewrite Equation \ref{eq:acc} as:
\begin{equation}\label{eq:deltaacc}
    \vec{\tilde a}(\omega) = \frac{4\pi}{3}G\rho_0\left(2\vec{\xi_P}(\omega)-\vec{\xi_S}(\omega)\right),
\end{equation}
where $\vec\xi_{P,S}$ is the seismic displacement vector in the frequency domain for the compressional (P) and shear (S) waves, respectively, and $\omega$ is the angular frequency. The derivation of Equation \ref{eq:deltaacc} can be found in Section 3.3 of \cite{harms} under the assumption that the underground seismic spectra are representative of body waves only and surface waves are negligible at a depth of a few hundred meters. Converting it into displacement, we have:
\begin{equation}\label{eq:deltacc2}
    \sqrt{S_{x_1-x_2}} = \frac{\tilde a(\omega)}{\omega^2} = \frac{4\pi}{3}\frac{G\rho_0}{\omega ^2}\left(2\vec{\xi_P}(\omega)-\vec{\xi_S}(\omega)\right).
\end{equation}
Moreover, we note that the PSD of the measured ground motion $S(\xi, \omega)$ can be expressed as the superposition of P and S waves contributions by defining a mixing parameter $p$: $S(\xi_P, \omega) = pS(\xi, \omega)$ and $S(\xi_S, \omega) = (1-p)S(\xi, \omega)$. As a consequence, assuming the lack of coherence which cancels the cross term,
\begin{equation}
    S_{2{\xi_P}-{\xi_S}} = 4S(\xi_P, \omega) + S(\xi_S, \omega) = (1+3p)S(\xi, \omega).
\end{equation}
From Equation \ref{eq:deltacc2}, we get
\begin{equation}
    2\vec{\xi_P}(\omega)-\vec{\xi_S}(\omega) = \sqrt {S_{2{\xi_P}-{\xi_S}}}=\sqrt{(1+3p)S(\xi, \omega)}.
\end{equation}
Assuming that the energy partition spreads uniformly across the degrees of freedom of body waves (1 for P waves, 2 for S waves) we set p = 1/3. Defining $\tilde x(\omega) = \sqrt{S(\xi, \omega)}$ as the amplitude spectral density of measured ground motion and plugging everything in Equations \ref{eq:deltacc2} and \ref{eq:deltaacc}, we get
\begin{equation}\label{eq:NNest}
    \tilde{h}_{\rm{NN}}(f) = \frac{4\pi}{3}G\rho_0\frac{2\sqrt{2}}{L}\frac{1}{(2\pi f)^2}\tilde{x}(f).
\end{equation}

The assumption of $p=1/3$ is done because the composition of the seismic fields is not known for the sites of interest, and so an ad hoc assumption was made \cite{alloccaetal2021, janssens2022, janssens2024, Badaracco:2021prd, harms2022, digiovanni2025} that displacement is equally distributed among the three degrees of freedom of seismic waves. The actual value of $p$ might be different for each site and depends on the types of seismic sources, their distance and the amount of scattering that affects the a wavefield. 

It is worth pointing out that Equation \ref{eq:NNest} is based on several assumptions that, if on one side make it a fast, reliable and widely used \cite{alloccaetal2021, digiovanni2025} method to estimate NN given realistic seismic noise levels acquired at each candidate site, on the other hand require the development of more robust and thorough models which take into account the geological substrate properties of the candidate sites. Studies aimed at the development of more accurate models for the composition of the wave field and for the properties of underground NN are currently underway within the ET collaboration (see, e.g., Ref. \cite{schillings2026numericalframeworknewtoniannoiseestimation}).

\section{The experience with Virgo}\label{sec:experience}

The Virgo interferometer offers valuable experience for understanding the impact of environmental disturbances on ground-based GW detectors \cite{Fiori_2026}. Since the beginning of its operations, Virgo has been characterized by an intricate environmental noise scenario, resulting from the complex interaction between natural seismicity, anthropogenic activity, local infrastructures and the detector itself. This interplay of environmental noise sources often proves as a major challenge, since the establishment of a direct causal relation between specific environmental noise sources and their impact
on the detector sensitivity is made complex by the coexistence
of multiple noise sources, the complexity of the coupling mechanisms and the non-stationarity of both the environmental disturbances and the interferometer response. Generally speaking, the experience acquired during Virgo commissioning and observing runs demonstrated that environmental noise is not only a limitation for the detector sensitivity at low frequency, but also a major factor affecting the duty cycle and detector stability.

Limiting the discussion to environmental disturbances related to noise sources of natural origin only, during the O3 observing run \cite{o3noise} there has been a lock loss every $\simeq$10 days due to an earthquake. Additional lock losses caused by earthquakes were found during periods in which the detector was not in science mode. To the purpose of preventing lock losses from earthquakes, the Virgo detector has an early warning system (Seismon) to get earthquake alerts from the low-latency US Geological Survey stream, which provides seismic wave arrival times and amplitude estimation at detector location. This system is interfaced with the Virgo data acquisition and control system to put the detector in a safe mode and prevent the lock loss.
However, on some occasions, this system may fail. This typically happens for either strong earthquakes far away or weak events nearby \cite{o3noise}. In the former case the amplitude of seismic waves can be strong enough to breach the detector's safe mode; for the latter, nearby events can be either too close to trigger early warnings or too weak to be reported to Seismon \cite{o3noise}.

Wind also plays a role in the operations of Virgo. For example, a consequence of high-wind conditions is the need for the Virgo global control system to use larger corrections to keep the instrument at its nominal working point. The larger the corrections are, the more the detector is vulnerable to additional disturbances that could make the corrections saturate and lead to an almost immediate control loss \cite{o3noise}. Overall, during the O3 science run it was found that the sensitivity of the detector is mostly unaffected up to wind speeds of $\simeq$20 - 25 km/h. At higher wind speeds the BNS range of the detector decreases up to $\simeq$4 Mpc for a wind speed of 50 km/h or above \cite{o3noise}. In any case, this variation is only about 10\% of the nominal BNS range values during O3, meaning that the detector is quite robust against wind. Studies to quantify the effect of wind on the O4 observing run are still in progress.

\begin{figure}[t]
\centering
\includegraphics[width=9.0 cm]{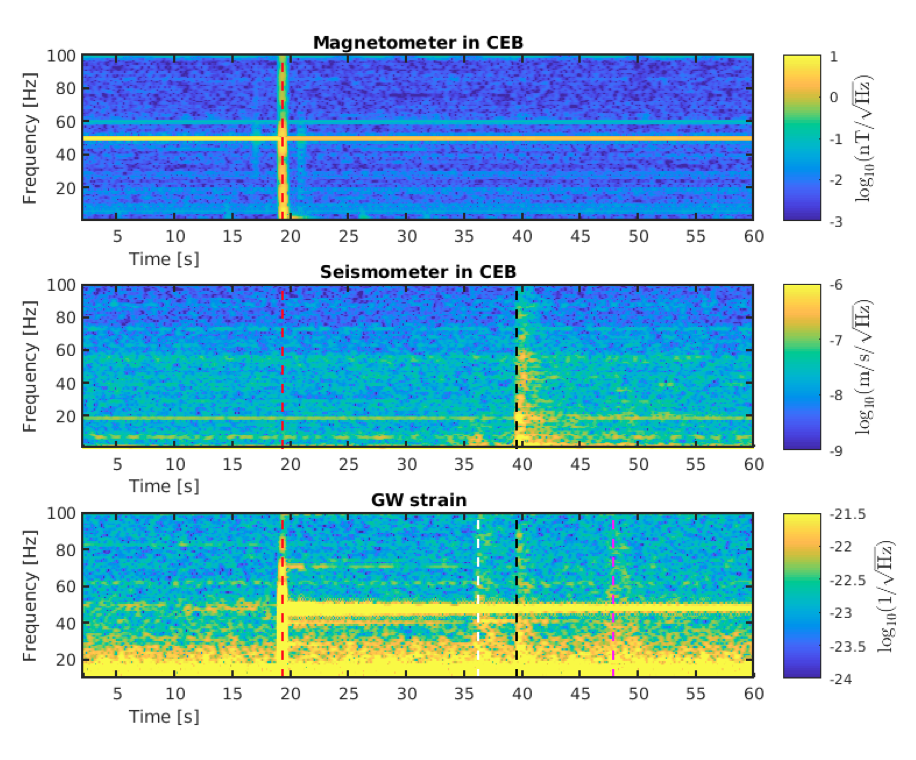}
\caption{Impact on the Virgo environment and detector of a lightning strike which occurred 6 to 10 km away from Virgo buidings on November 15, 2019 at 23:25:51 UTC. (Top) A prompt magnetic transient is detected by magnetometers at the time of the event, marked by the red vertical line. (Middle) A few tens of seconds later, a seismic (and acoustic, not shown) transient is detected in the central experimental area, marked by the black vertical line. The bottom spectrogram shows the reconstructed GW strain during the same time interval. The red vertical line marks the lightning strike occurence, the black, magenta and white vertical lines mark the occurrence of seismic transients detected in the CEB, NEB and WEB, respectively. This figure is from \cite{o3noise}. \label{fig:lampo}}
\end{figure}

Overall, environmental disturbances are often not limited to a single channel, but they manifest themselves through different paths. Figure \ref{fig:lampo} shows an interesting example of a glitch visible in different channels as well as in the reconstructed GW strain channel of the detector: a lightning strike occurring a few km away from the detector caused a glitch in both magnetic and seismic noise; this glitch propagated to the reconstructed GW strain during the same time interval.

One of the most relevant lessons learned from Virgo and current GW detectors concerns the importance of detector characterization activities and continuous environmental monitoring. In particular, the Virgo detector is equipped with an extensive Physical Environmental Monitoring (PEM) system composed of seismometers, accelerometers, microphones, magnetometers, weather stations and auxiliary sensors distributed both inside and outside the experimental areas.\cite{cvse,o3noise} The goal of the PEM is to identify environmental disturbances, reconstruct their coupling with the interferometer output and monitor the evolution of environmental conditions in time to maximize the scientific reach of the detector. A significant fraction of the environmental noise affecting Virgo originates from anthropogenic activities surrounding the detector site. Road traffic, railways, industrial activities, wind turbines and planes contribute to generating seismic and magnetic disturbances observable in the interferometer data.\cite{acernese2004properties, virgo2006, saccorotti, poli, piccinini, o3noise} On several occasions, transient disturbances generated by trains or local infrastructures were observed to couple to the interferometer through scattered light or suspension resonances, producing glitches and excess noise in the GW strain channel.\cite{Fiori2020,o3noise}
\begin{figure}[t]
\centering
\includegraphics[width=9.0 cm]{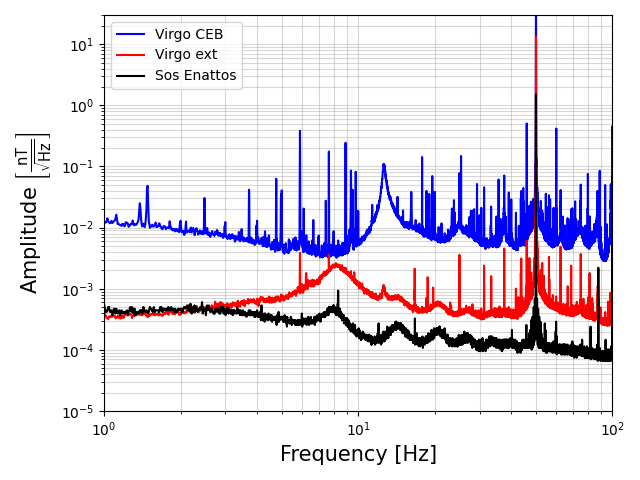}
\caption{Comparison of the amplitude spectral densities of indoor (red curve) and outdoor (blue curve) magnetometers at Virgo and at Sos Enattos mine in Sardinia 110 m underground (black curve). The red and black lines show evidence of Schumann resonances peaked at approximately 8, 14, 21, 27 and 33 Hz.\label{fig:magcomp}}
\end{figure} 

The Virgo experience also highlighted the importance of self-inflicted noise, namely environmental disturbances generated by the detector's infrastructure itself.\cite{o3noise,infra, Fiori_2026} Vacuum pumps, cooling systems, ventilation systems, transformers, electronics racks and power distribution systems can generate seismic, acoustic and magnetic noise that propagate through the experimental infrastructure and couples with the detector subsystems. In some frequency bands, self-generated disturbances were found to dominate over external environmental noise \cite{o3noise}. 

An example is shown in Figure \ref{fig:magcomp}: at the Virgo site, the external magnetic environment is much quieter than inside the experimental areas, which are characterized by stray magnetic fields radiated by electric loads and cables where large currents are circulating. Comparison against a magnetometer located at the ET candidate site at 110 m undeground inside the Sos Enattos mine in Sardinia is also shown. The most intense spectral noise features are narrow lines at the 50 Hz electric mains frequency and its odd harmonics. The RMS amplitude of the 50 Hz line measured at Virgo is of the order of 0.1 nT in the external location, while it is at least 50 times larger in any inside location. This demonstrates that the environmental compatibility of each subsystem must be considered as part of the detector design itself to avoid self inflicted noise, rather than as a secondary engineering problem.
\begin{figure}[t]
\centering
\includegraphics[width=9.0 cm]{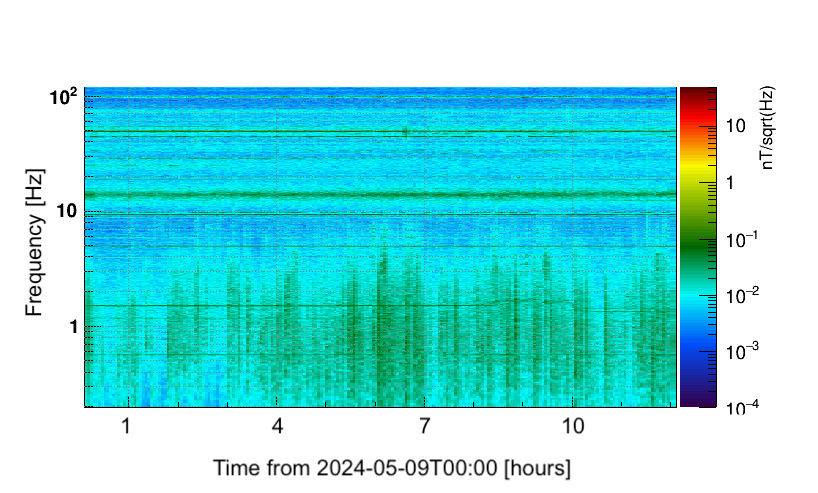}
\caption{Example of a time frequency plot obtained from a magnetometer located in one of the Virgo's experimental areas. Comparison against Figure \ref{fig:magnoise}.a reveals a much more complex environment, especially in the low frequency range, with respect to a site without significant infrastructures in the neighborhood. \label{fig:magvirgo}}
\end{figure} 

 Overall, magnetic disturbances produced by power lines, electrical infrastructures and moving ferromagnetic objects were observed to couple with the interferometer through coil-magnet actuators and other magnetic components. In Figure \ref{fig:magvirgo} we show the time frequency plot from a magnetometer located in one of the experimental areas of Virgo; comparison against Figure \ref{fig:magnoise}.a reveals a much more complex environment, mainly related to the presence of the infrastructure. The analysis of magnetic injections and coupling functions demonstrated that even weak environmental magnetic fields can generate measurable disturbances in the strain channel, even though they do not affect the sensitivity of current detectors up to a BNS range of 100 Mpc.\cite{cirone} Moreover, correlations between magnetic transients and glitches in GW data were observed during detector operations \cite{o3noise}. These studies motivated the development of dedicated magnetic monitoring strategies and mitigation procedures for ET and for the selection of a site with as few external infrastructures as possible, in order to have a magnetic background which is mainly related to natural sources (Schumann resonances, geomagnetic pulsations, solar storms) and to offer a better control and characterization of magnetic noise sources from the detector infrastructure.

Acoustic noise also played a significant role during Virgo operations. Air handling systems, fans and human activities inside the experimental halls generated acoustic disturbances capable of coupling with optical benches, suspended components and electronics.\cite{infra} Acoustic disturbances were frequently associated with scattered light phenomena and with the excitation of mechanical resonances of auxiliary components. As a consequence, the reduction of acoustic coupling required both mechanical mitigation strategies and strict operational procedures during detector operation. As an example, during the O4 observing run, a significant effort has been put to reduce the low-frequency noise produced by the
Virgo heating, ventilation and air conditioning (HVAC) systems \cite{Fiori_2026} and reduce the risks for future upgrades of the detector. HVAC related disturbances, namely from both rotating machinery and air and water circulation systems, can reach the experimental areas through several transmission paths, including propagation through the ventilation ducts, structure-borne vibrations through mechanical connections and building structures and pressure fluctuations transmitted through water circuits. However, focused actions allowed to achieve significant mitigation of both acoustic and seismic noise: changes in fan shapes at the Virgo terminal halls greatly reduces acoustic and seismic noise below 70 Hz; the adoption of commercial solutions such as damped springs and high-density damping materials reduced seismic noise that originated from the direct transmission of airborne noise from the centrifugal fan through the duct network; vibrations water-borne pressure fluctuations from circulation pumps and propagating through the pipe network were mitigated by reducing pumps speed and isolating unused pipe branches \cite{Fiori_2026}.

Overall, this HVAC-targeted mitigation campaign demonstrated that substantial reductions in infrastructure induced noise can be obtained through a combination of source control and damping of transmission paths. These achievements emphasize how coupling mechanisms and mitigation systems for infrastructure services should be carefully accounted for during the design and operation of GW observatories, rather than treating systems such as HVAC solely as conventional building services. This also demonstrates that future detectors require an intrinsic low-noise facility design, improved mechanical isolations and optimized ventilation systems to grant both a livable and safe underground laboratory and a sustainable infrastructure for the scientific potential of ET.

Finally, Virgo, as well as LIGO, identified another potential threat to ground based GW detectors: wind farms \cite{ligowind,saccorotti}. The towers of the wind turbines vibrate with several bending and torsional modes, mainly due to the wind pressure and blade motion. Vibrational modes propagate to the ground, translate into seismic noise through surface waves and are visible as peaks in seismic spectra. If one of these vibrational modes happens to coincide with, e.g., a resonance of the seismic isolation system, like those shown in Figure \ref{fig:TF}, it could significantly increase the contribution of the resonance to the r.m.s. of the noise at that frequency and require dedicated mitigation measures. One  of the most critical aspect of seismic noise from wind farms derives from the fact that damping of seismic noise with distance shows a large variability with the geological and topographical characteristics of the site, with the layout of the wind farms and with the model of the wind turbines, often resulting in complicated patterns \cite{saccorotti}. The intensity of the peaks is also closely related to wind conditions \cite{diaferia25}. Finally, vibrational modes also propagate through the air via the emission of low frequency pressure waves (infrasound) which couple locally into the ground \cite{saccorotti}. Therefore, the impact of the noise produced by present or future wind farms is strongly site-dependent and must be carefully evaluated with dedicated geophysical studies. Any plan of building new wind farms, or expand the existing ones, in the vicinity of GW detectors must be carefully reviewed by the scientific community before approval or stopped through the establishment of exclusion zones by local governments.

In general, from site characterization studies for ET and environmental monitoring in Virgo, we learned that environmental disturbances vary on daily, seasonal and meteorological timescales.\cite{digiovanni2023,o3noise} Wind conditions, sea activity, temperature variations and human activities continuously modify the environmental conditions surrounding the detector. Therefore, detector characterization cannot be considered as a static problem, but rather as a continuous monitoring and mitigation effort that accompanies the entire lifetime of the experiment. As a consequence, the mitigation of environmental noise requires a global approach involving detector design, infrastructure planning and operational procedures. These lessons constitute one of the most important starting points for the design and operation of ET, which will feature an infrastructure design that  not only shields against external noise but also rigorously minimizes the emission of acoustic, vibrational, and electromagnetic noise from its own operational plant.

\section{Controlling the environment}\label{sec:control}

The experience acquired with current GW detectors demonstrated that environmental noise mitigation cannot rely exclusively on seismic isolation systems or on post-processing subtraction techniques. Instead, the environmental sustainability of the entire experimental infrastructure must be addressed from the earliest stages of the detector design. This aspect is particularly relevant for ET, whose low-frequency sensitivity target makes the detector substantially more susceptible to environmental disturbances than current interferometers. All of this combined is pushing the scientific community to improve not only our knowledge of possible noise sources, but also to design a sustainable infrastructure with the smallest possible impact on the sensitivity of future GW detectors. 

Considering our current knowledge of GW detectors and noise sources, we can list cryogenic systems, ventilation systems, cooling plants, power distribution networks, pumps and auxiliary facilities as potential sources of seismic, acoustic and magnetic disturbances. Therefore, the construction of ET requires a careful balance between the technological needs of the infrastructure and the environmental compatibility of the experimental site. Moreover, infrastructures excavated underground may themselves become sources of vibrations, acoustic resonances and magnetic perturbations. Consequently, underground construction must be complemented by careful environmental engineering. 

For example, the KAGRA detector provides a unique testbed for the operations of a GW detector in an underground setting. KAGRA is situated at a depth of over 200 meters in the Kamioka mine (Japan). This setting provides a natural shield against surface seismic disturbances, effectively suppressing atmospheric pressure variations and exponentially attenuating NN from surface Rayleigh waves \cite{Badaracco:2021prd, Yokozawa:2023icrc}. To be more specific, in the frequency band above 1\,Hz, the ground vibration level is suppressed by a factor $\mathcal{O}(10^2)$). Infrastructure components such as ventilation systems and cryogenic machinery couple primarily via localized acoustic fields rather than generating long-range seismic waves, restricting infrastructure disturbances to the immediate vicinity of the utility stations \cite{Badaracco:2021prd}. However, the complex infrastructure itself introduces distinct noise challenges. For instance, the KAGRA underground infrastructure needs a system of water pipes to collect humidity and water percolating from the rocks above. This means that the internal fluid dynamics within nearby water pipelines can eventually contribute to NN and will be a potential issue for ET, which will require a similar system to collect underground water\cite{Yokozawa:2023icrc}. In addition to that, essential machinery can excite unexpected mechanical resonances, such as a 700 Hz vibration identified on the Output Mode Cleaner vacuum chamber framework \cite{Yokozawa:2023icrc}. 

\subsection{The case of wind farms}

As already mentioned in Section \ref{sec:experience}, wind farms in the neighborhood of GW experiments are considered a potential risk for the operations of the detectors. To this purpose, the ET community is carefully evaluating and characterizing the seismic noise signatures, both on surface and underground, of these facilities close to the ET candidate sites. A notable example of these efforts is the WINES experiment \cite{diaferia25}, which installed an array of seismometers between the Sardinia ET candidate site and the Buddusò wind farm, some 17 km away from the site. The combination of surface and borehole seismic measurements with the logs of the wind turbines about wind speed and blade rotational speed allowed to carefully monitor the noise signature of the wind farm at the ET candidate site. Consistently with previous studies at the Virgo \cite{saccorotti} and LIGO \cite{ligowind} sites, the decay of the spectral amplitudes of the frequency peaks related to the wind turbines has been found to follow a $r^{-\alpha}$ trend, with $r$ being the distance from the wind farm and $\alpha$ depending on frequency and on the blade rotation speed. On top of that, it was found that when the wind park approaches its maximum regime, it generates a seismic noise signature that can be recorded even by the borehole sensors at a depth of 250 m, more than 17 km away.

These results, in connection to similar studies conducted at the EMR and Lausitz sites, pushed local governments to issue restrictive measures to stop plans for expanding the existing wind farms or to build new ones. Simultaneously, some research units within the ET collaboration are evaluating engineering solutions aimed to reduce the vibrations generated by wind turbines. These proposal include, e.g., the installation of mass dampers inside the turbines' masts, larger foundations \cite{boxberg} or the adoption wooden masts \cite{wood}.

\subsection{Environmental noise mitigation strategies}
The development of appropriate noise mitigation strategies is one of the key aspects taken into account in the design of ET-LF, which, as we have seen, requires a comprehensive mitigation of environmental noise sources. Since these are closely tied to the features of each ET candidate site, the achievable performance, robustness and cost of mitigation strategies will be strongly site dependent, as they are influenced by geological conditions and long-term environmental stability. The process of monitoring and mitigating environmental noise relies on a combination of advanced sensor technologies, optimized infrastructure design and data-driven noise-cancellation methods.

As far as NN is concerned, the most commonly accepted method for mitigation and cancellation is the use of the Wiener filter to remove it from the data stream of the detector (see, e.g.,\cite{harms15} and references therein for the classic implementation of the static Wiener filter, or, for a newly proposed adaptive filtering method, see \cite{Koley2024}). Even if they are not limited by NN, the Wiener filter method has been tested in current detectors and relies on the deployment of dense arrays of environmental sensors to monitor the seismic and atmospheric fields which surround the test masses \cite{badaracco2019, Badaracco2020}. Currently, the use of neural networks is also being investigated to support and improve NN mitigation strategies \cite{vanBeveren_2023, kelleter2026nnnnneuralnetworksnewtonian}.

The sensors used by mitigation systems include seismometers, deployed at the surface or in boreholes, tiltmeters, strain sensors, infrasound microphones and atmospheric probes \cite{harms2022, Bader_2022}. The deployment of extensive arrays of environmental sensors is needed to maximize the performance of mitigation systems that rely on the accurate knowledge of the environment, such as the composition of the seismic wave field, the relative contributions of surface waves and body waves, as well as on their spatial coherence \cite{Andric2020}. Other unpredictable variations caused by, e.g., anthropic activity, can alter the field correlation structure between sensors and test-mass motion, requiring adaptive filtering strategies and periodic re-optimization of sensor layouts \cite{badaracco2019, Badaracco2020, Koley2024, Schillings_2025}. It is clear that a possible limitation to noise cancellation strategies is given by the number of sensors needed to achieve acceptable mitigation factors. Recent studies suggest that achieving the required NN suppression factor (from 3 to 6) may necessitate of up to 15-20 seismometers per test mass \cite{badaracco2019, Badaracco2020}. Taking into account that, in the case of the triangular configuration, ET-LF will have 12 test masses, we  could have a variable number, between 100 and 200, of seismometers deployed in boreholes at variable depths down to several hundred meters \cite{badaracco2019, Badaracco2020}. Other works, such as Ref. \cite{kelleter2026nnnnneuralnetworksnewtonian}, also show that achieving NN suppression factors significantly more than 10 will require up to few hundred sensors per test mass, whilst delivering figures consistent with \cite{badaracco2019, Badaracco2020} for suppression factors from 3 to 6.

However, given the swift evolution of NN mitigation strategies, these figures should not be taken as definitive and are expected change with the development of new technologies. As a consequence, \textbf{it should now be clear that} large-scale NN mitigation systems represent a substantial \textbf{research and development,} financial and infrastructural investment that has to be planned with great advance. To this purpose, cost effective strategies for sensor deployment \cite{schillings2026b} and computational infrastructure \cite{Schillings_2025} and plans for the development of new and more economic technologies for the sensors, such as distributed acoustic sensing \cite{Ophardt_2026}, are constantly being evaluated inside the ET collaboration.

On the other hand, mitigation strategies for magnetic noise are still in a preliminary phase. In fact, being the reduction of the magnetic susceptibility of the detector closely tied to its design, reduction strategies for magnetic noise are still in a preliminary phase. These strategies will be aimed at studying the magnetic properties of suspension components, actuator design and shielding of sensitive subsystems which can significantly reduce the impact of magnetic noise on the detector.
    
Mitigation strategies can also be limited by several other factors. The scattering of seismic waves and the composition of the seismic field can pose a threat to NN subtraction performance \cite{Andric2020}. This limitation arises from the fact that the optimal positions of the underground sensors for NN mitigation has to be decided in advance and designed according to specific conditions of the seismic field. Any unpredicted variation in the seismic field, as well as time-varying noise sources of human origin, may hinder the performance of NN suppression systems. In conclusion, environmental-noise mitigation technologies and strategies have to find the appropriate compromise between site quality, achievable low-frequency sensitivity and cost.

\subsection{Impact of correlated Newtonian noise in co-located detectors}
Given the nature of the triangle configuration, where the test masses of different detectors are expected to be located at a distance of a few hundreds meters \cite{ET2020} from each other, Newtonian noise can become a correlated noise source between the detectors, if not properly mitigated. Refs. \cite{janssens2022, janssens2024} delivered a first estimate of the impact of correlated Newtonian noise for the ET triangular configuration in the case of searches for the isotropic gravitational-wave back-
ground. In this case, the search is based on the correlation between interferometers. As a consequence, a spurious correlation can lead to a significant decrease in sensitivity. In particular, Ref. \cite{janssens2024} found that the search for an isotropic gravitational wave background will be limited by correlated Newtonian noise in case of a co-located detector configuration, as coherence drops significantly only for test masses separated by at least 2.5 km.

Moreover, in other cases, the combinations of the signals from each interferometer can be used used to reconstruct the GW parameters. If the noises from the three detectors are independent, they can be added in quadrature to reach the target sensitivity, whereas, if they are correlated, we cannot expect a gain in signal-to-noise ratio by combining them.

It is also worth pointing out that Newtonian noise generates a displacement noise which is inversely proportional to the arm length of the detector \cite{CE4}. As a consequence, the 2L configuration can be favored by a factor of 1.5 over the triangle, due to the difference in arm length. The importance of Newtonian noise subtraction is also higher in the triangle with respect to the 2L detectors, given that any residual NN noise might be correlated between the three LF detectors, due to the proximity of test masses.

\subsection{Impact on duty cycle}
As demonstrated by the operations of current detectors \cite{gwtc1, gwtc2, GWTC3, gwtc4, gwtc5}, the achievement of high duty cycle rates is of vital importance to meet the science goals of the detectors. Because of this, the peculiar configuration of the proposed ET triangular detector, foreseeing six co-located detectors, presents higher risks for what concerns the hypothetical impact on the duty cycle of coincident environmental disturbances and maintenance activities. 

To solve this problem, a recent study \cite{Negri:2026clm} proposed a rotating maintenance approach for the triangular detector. This proposal consists in planning a maintenance schedule to work on a single detector while keeping the other two operational. Ref. \cite{Negri:2026clm} tried to validate this approach by showing that the impact of scheduled down times on the observations of binary black hole mergers is lower than in the case of a 2L network. This finding relies on the fact that in the triangular detector case there will be 2 operational detectors most of the time, contrary to the 2L case. However, the work presented in Ref. \cite{Negri:2026clm} relies on the quantitative and hypothetical belief that maintenance activities will not generate any excess noise or glitches that can affect the operations of the other two detectors in the network. As a matter of fact, from the operations of current detectors, we know that maintenance activities involve complex tasks which often require the use of machinery with several people moving around on the site. Considering the impact of human activities on background seismic noise \cite{digiovannietal2021, digiovanni2023, digiovanni2025, saccorotti23} and the possible occurrence of magnetic glitches appearing due to the operations of power tools and other devices \cite{cirone, mnvirgo}, the presumption of maintenance activities not impacting co-located detectors seems less likely. In any case, adequate maintenance schedules to reduce down times in the triangular configuration do not take into account coincident environmental noise transients that may affect all three detectors at the same time. As we have seen in Section \ref{sec:experience}, the local environment has a significant impact on the operations of a GW detector (e.g. Figure \ref{fig:lampo}). Operations of three detectors at the same location puts the observations at the risk of being subject by the same environmental glitches affecting all three detectors at the same time.

\subsection{Long-term preservation of the site}
In the case of ET, strict constraints on heavy traffic, industrial activities, transportation infrastructures and construction sites in the surroundings of the detector may be necessary to preserve the low-frequency sensitivity target and to grant a suitable environment for the entire life cycle of the detectors. This implies that the long-term scientific exploitation of ET requires preserving the quality of the surrounding area over several decades. This introduces the need for dedicated environmental protection policies around the detector site. In fact, a site which is compliant with the environmental requirements of ET as of 2026, it may not be so in 50 years from now. Industrial expansion, new transportation infrastructures, wind farms and other anthropogenic activities in the proximity of the detector should be limited if not banned.\cite{amann,digiovannietal2021,digiovanni2023} Therefore, environmental quietness should be regarded as a strategic scientific resource to preserve throughout the operational lifetime of the observatory.

It is worth pointing out that the preservation of the environment to pursue scientific objectives of an experiment is not a novelty. Over the last century, this topic has been a matter of concern notably for astronomers who operate telescopes \cite{West1985,Goddi2025} as well as for research facilities that operate equipment sensitive to environmental perturbations. For instance, the geophysical Black Forest Observatory and the Moxa geodynamics observatory, both in Germany, are the center of exclusion zones aimed at avoiding an excess of ground motion from civil infrastructures. Moreover, the operations of instruments sensitive to electromagnetic interferences led some local governments to establish radio free zones around specific research facilities \cite{FCC11745_NRQZ_1958}, whereas the growth of satellite constellations, such as Starlink, is posing a serious threat to radio astronomy \cite{Borlaff2025_Megaconstellations} and demands for global actions. On top of that, with fundamental physics experiments constantly increasing their sensitivity to achieve groundbreaking results and increasing in scale, the quality of the noise field of a site hosting an experiment has become a matter of growing concern in the broader scientific community. In this sense, the Einstein Telescope will be at the forefront of site-protection policies, requiring unprecedented precautions to preserve the quality of the noise environment of its host site.

\subsection{Climate change and long-term environmental stability}
\label{sec:climate}
We have seen that the environmental conditions at the site of a gravitational-wave observatory are not necessarily stationary over the multi-decade lifetime of the experiment, mainly due to the urban and economic development on surface. This non-stationarity can be controlled by appropriate measures that limit surface activities. On the other hand, there are other possible non stationarities which cannot be controlled but can pose a potential threat to the quality of the noise environment. In fact, climate change may affect several environmental parameters relevant to the operation of ET. Changes in precipitation patterns, temperature, wind regimes and the frequency of extreme weather events may modify the environmental conditions surrounding and within the underground infrastructure. In particular, variations in precipitation and groundwater circulation could affect the hydrological conditions of the host rock, potentially modifying local ground deformation, seismic activity and the operation of water-management systems. Similarly, changes in temperature and atmospheric conditions may affect air circulation, pressure fluctuations and the performance of ventilation and cooling systems.

Extreme weather events represent another potential challenge. Strong winds, intense precipitation and severe atmospheric disturbances can generate transient seismic and acoustic disturbances and can affect the operation of surface infrastructure, power systems and access facilities. Even when the interferometer itself is protected by the underground location, disturbances generated at the surface may propagate through the infrastructure or affect auxiliary systems. The increasing frequency or intensity of extreme events could therefore have consequences not only for detector sensitivity, but also for detector availability and operational procedures.

From this perspective, environmental monitoring should not be considered only as a tool for characterizing the noise background during commissioning and observing runs. Long-term monitoring of seismic, atmospheric, hydrological and meteorological parameters can provide the information required to identify secular changes in the environmental conditions of the observatory. Such measurements could also help distinguish intrinsic changes in the detector from changes in the surrounding environment and provide a baseline against which future environmental evolution can be assessed.

\section{Site dependent noise and the impact on the sensitivity curve}\label{sec:dependent}

It should now be clear that, unlike current detectors whose sensitivity below $\sim 20\,\rm Hz$ is limited by their intrinsic design, ET aims at exploring a frequency region in which environmental disturbances become one of the primary limiting factors. Consequently, the characteristics of the seismic and environmental field surrounding the detector could directly influence the achievable sensitivity curve and the scientific capabilities of the observatory.

The seismic field measured at candidate ET sites shows significant differences depending on local geology, depth, proximity to the sea, meteorological conditions and anthropogenic activity (Figure \ref{fig:spectracomp}).\cite{digiovannietal2021,digiovanni2023,saccorotti23,Bader_2022} In particular, underground measurements performed at the Sardinia, EMR and Saxony candidate sites revealed a substantial variability in the seismic spectrum above 1\,Hz, precisely in the frequency band most relevant for ET-LF. The Sardinia site, characterized by extremely low anthropogenic activity and favorable geological conditions, exhibits seismic spectra approaching the NLNM in the frequency range relevant for ET.\cite{digiovanni2023} On the other hand, sites located in densely populated or industrialized regions show higher levels of anthropogenic seismic noise.
\begin{figure}[t]
\centering
\includegraphics[width=7.0 cm]{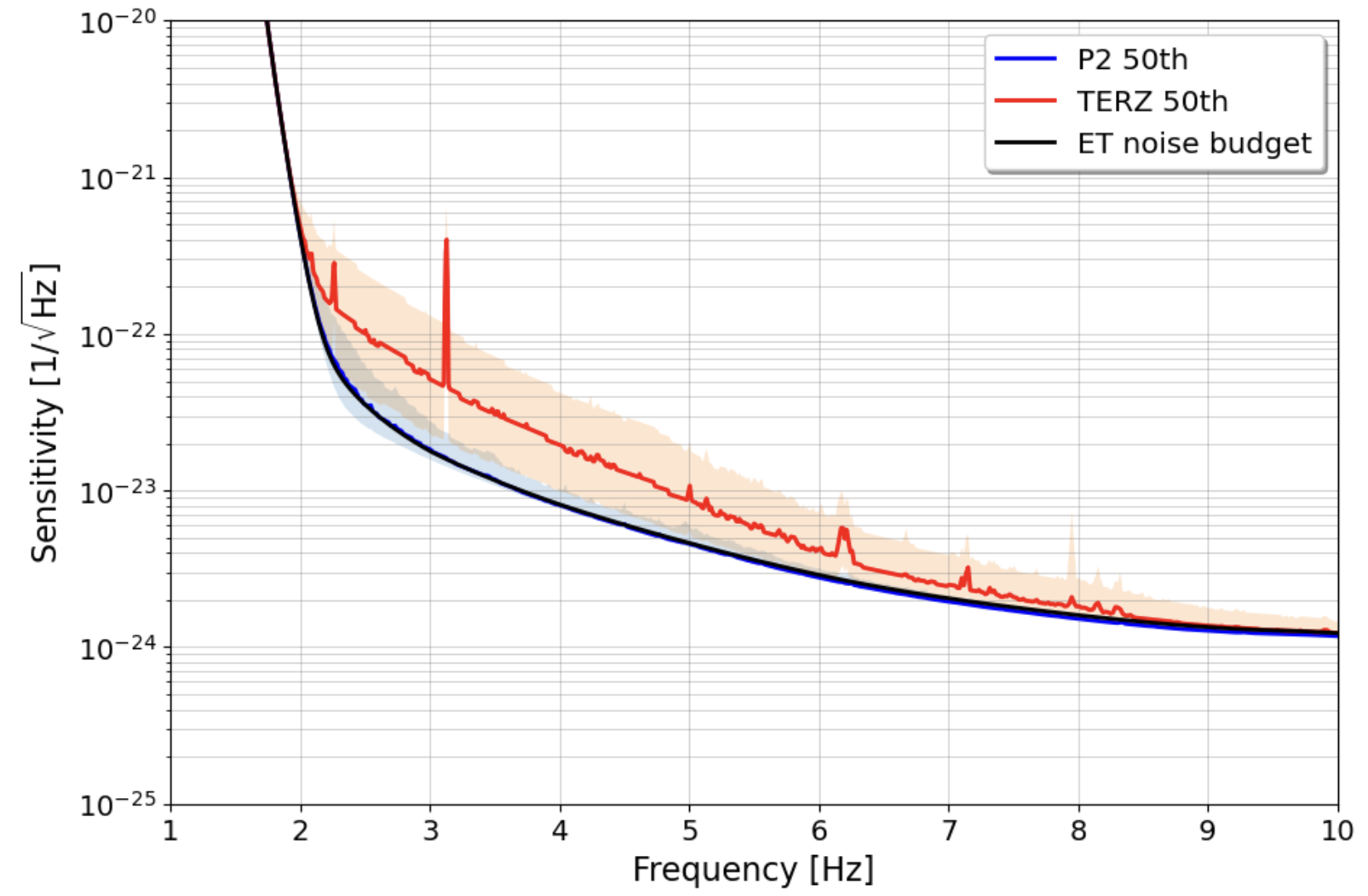}
\caption{Sensitivity curves for the two candidate sites, reporting the percentile (50th in solid line while 10th and 90th the lower and upper limit of the colored band) for TERZ (in red) and P2 (in blue), compared with the ET design sensitivity (black). This plot is taken from \cite{digiovanni2025}. \label{fig:cmp}}
\end{figure}  
These differences are not merely local geophysical features but could have direct consequences on the detector performance. As an example, Figure \ref{fig:cmp} shows, using the analytical approximation of Equation \ref{eq:NNest} for NN, a comparison of the impact of local noise on the ET sensitivity curve at the Sardinia and EMR candidate sites. In this case, we decided to not include any mitigation factor for NN in the noise budget. This was done not only because, at the moment, it is hard to make a forecast on a realistic value of this mitigation factor (any choice would be arbitrary, see Section \ref{sec:control}), but also to outline, once again, how noise mitigation strategies will have to be designed according to the unique features of environmental noise recorded at each site and will need to focus on specific frequencies. There will not be a unique method or design which can be applied regardless of the location of the detector.

In particular, using this approach, \cite{alloccaetal2021,digiovanni2025} demonstrated how the dependence of the ET sensitivity on local environmental conditions has important astrophysical implications (Figure \ref{fig:cmp}). The detectability of low-frequency inspiral signals from compact binaries strongly depends on the sensitivity achieved below $\sim 10\,\rm Hz$. This frequency range is crucial for early warning generation in binary neutron star mergers and for the observation of intermediate mass black hole inspirals. A degradation of the low-frequency sensitivity caused by high environmental noise levels may reduce the duration of signals observable in band, the achievable signal-to-noise ratio and the effectiveness of multimessenger follow-up observations. It is also worth pointing out that Figure \ref{fig:cmp} includes only the NN contribution from the seismic field. The contribution from atmospheric NN as well as magnetic noise and pure seismic noise are still being investigated to produce a realistic assessment of the foreseen ET sensitivity at low frequency.

As well as NN, the ET collaboration is also in the process of assessing the site dependent impact of magnetic and acoustic noise. The goal is to deliver appropriate transfer functions which will project each noise source to the sensitivity of ET. The main issue in this sense is the current lack of a credible projection, since it cannot be based on an actual infrastructure for ET and have to rely on the experience with Virgo.

The site dependence of environmental noise is also expected to have an impact on detector operations and commissioning. A noisier environment implies stricter requirements for seismic isolation systems, more complex noise subtraction procedures and increased infrastructure mitigation efforts. Consequently, the characteristics of the site influence not only the scientific output of the detector but also the complexity and cost of its operation. For these reasons, site characterization studies cannot be regarded as complementary investigations performed after detector design. Instead, they represent a fundamental component of the ET project and directly contribute to defining the achievable detector sensitivity. In this context, the environmental properties of the host site become part of the detector itself.

\section{Conclusions}\label{sec:conclusions}
In this paper we discussed the role of environmental noise and environmental conditions in future GW interferometers, with a particular focus on ET. The extension of the observational bandwidth down to $\sim 2,\mathrm{Hz}$ and the substantial improvement in sensitivity with respect to current detectors introduce unprecedented requirements on the environmental conditions surrounding the observatory. In the low-frequency range, the experience with current detectors demonstrate that environmental disturbances cannot be regarded as secondary effects, but may directly influence the achievable sensitivity, detector stability and scientific performance of the experiment. Given its sharp increase in sensitivity, these noise sources will play an even more important role in ET.

In this work, We reviewed the main classes of environmental disturbances which are expected to affect ET, including seismic, Newtonian, acoustic and magnetic noise, as well as disturbances associated with the detector infrastructure and anthropogenic activity. The experience accumulated with Virgo and other second-generation interferometers demonstrates that environmental disturbances can couple to the detector through a variety of mechanisms, including direct mechanical and electromagnetic coupling, scattered light, control-system interactions and nonlinear up-conversion processes. These observations also show that environmental noise is not a static property of the detector, but varies over daily, seasonal and meteorological timescales and can include both stationary backgrounds and transient disturbances. However, the use of appropriate mitigation strategies and careful design of the infrastructure can effectively reduce the impact of environmental noise sources.

A major conclusion emerging from the preparatory studies for ET is that environmental noise mitigation cannot be considered as independent from the design of the detector and its supporting infrastructure. The large underground infrastructure required by ET will provide significant attenuation of several external environmental disturbances, but may potentially introduce new potential noise sources through ventilation, cooling and cryogenic and other supporting systems and human activity. Consequently, the environmental compatibility of the infrastructure should be considered from the earliest stages of the design, with appropriate spatial separation, mechanical and acoustic isolation and operational procedures implemented as integral parts of the observatory.

The analysis also highlights the importance of the environmental characteristics of the host site. Seismic measurements at the candidate ET sites have demonstrated substantial differences in the environmental background, particularly in the frequency range relevant for the ET low-frequency sensitivity. These differences can propagate into the expected detector sensitivity and, consequently, into the observability of low-frequency gravitational-wave signals. In particular, the impact of seismic noise and Newtonian noise demonstrates that site characterization cannot be limited to the measurement of seismic amplitudes, but should also consider the composition and spatial properties of the environmental fields and their relevant coupling mechanisms. Similar considerations apply to acoustic and magnetic disturbances, whose impact will ultimately depend on the infrastructure and coupling functions associated with the final detector design.

The triangular configuration of ET introduces an additional aspect that is not fully captured by considering environmental disturbances as independent noise sources for each interferometer. Environmental fields generated by common natural or anthropogenic sources, as well as disturbances produced by shared infrastructure, may affect multiple interferometers simultaneously. Such correlations may have consequences for detector operation, data-quality procedures and coherent analyses. Understanding the spatial and temporal correlation properties of environmental disturbances will therefore be an important component of the environmental characterization and mitigation strategy of ET.

Environmental monitoring will consequently play a role that extends beyond conventional detector characterization. Distributed arrays of seismic, acoustic, magnetic, atmospheric and meteorological sensors will be required to characterize the environmental fields, identify transient disturbances and determine the relevant coupling functions. In the case of Newtonian noise, environmental sensing and reconstruction of the environmental noise field represent one of the few available approaches to mitigation. More generally, environmental monitoring will provide the information required to distinguish environmental disturbances from instrumental effects and may support real-time noise subtraction, data-quality assessment and detector control.

Finally, the environmental conditions relevant for ET should be considered on timescales comparable to the expected lifetime of the observatory. Site selection provides a characterization of the present environmental background, but the seismic, atmospheric, hydrological and anthropogenic environment may evolve over several decades. Changes in land use, transportation infrastructure, industrial activity, hydrological conditions and extreme weather events may modify the environmental conditions surrounding the detector. Climate change therefore represents an additional long-term challenge, not necessarily because its impact on ET sensitivity can currently be quantified, but because it may alter the environmental boundary conditions assumed during site selection and detector design. Long-term environmental monitoring and, where possible, preservation of the low-noise characteristics of the hosting site should therefore be regarded as part of the long-term operation strategy of the observatory.

Overall, the development of ET requires a transition from treating environmental disturbances as external noise sources to considering the environment as an integral component of the detector. The scientific performance of the observatory will depend on the combined optimization of the site, underground infrastructure, interferometer design, environmental monitoring and noise mitigation strategies. A multidisciplinary approach involving geophysics, atmospheric and environmental sciences, detector physics and infrastructure engineering will therefore be essential to fully exploit the low-frequency sensitivity of ET and to preserve its scientific capabilities throughout its operational lifetime.

\providecommand{\newblock}{}


\begin{thebibliography}{100}
\expandafter\ifx\csname url\endcsname\relax
  \def\url#1{{\tt #1}}\fi
\expandafter\ifx\csname urlprefix\endcsname\relax\def\urlprefix{URL }\fi
\providecommand{\eprint}[2][]{\url{#2}}

\bibitem{GW150914}
{LIGO Scientific Collaboration and Virgo Collaboration} 2016 {\em Phys. Rev.
  Lett.\/} {\bf 116}(6) 061102
  \urlprefix\url{https://link.aps.org/doi/10.1103/PhysRevLett.116.061102}

\bibitem{gwtc1}
{C}ollaboration L 2019 {\em Phys. Rev. X\/} {\bf 9}(3) 031040
  \urlprefix\url{https://link.aps.org/doi/10.1103/PhysRevX.9.031040}

\bibitem{gwtc2}
{LIGO Scientific Collaboration and Virgo Collaboration} 2021 {\em Physical
  Review X\/} {\bf 11} ISSN 2160-3308
  \urlprefix\url{http://dx.doi.org/10.1103/PhysRevX.11.021053}

\bibitem{GWTC3}
{LVK Collaboration} 2023 {\em Physical Review X\/} {\bf 13} ISSN 2160-3308
  \urlprefix\url{http://dx.doi.org/10.1103/PhysRevX.13.041039}

\bibitem{gwtc4}
{LVK Collaboration} 2025 {GWTC}-4.0: Updating the gravitational-wave transient
  catalog with observations from the first part of the fourth
  {LIGO}-{Virgo}-{KAGRA} observing run submitted 25 August 2025; revised 8
  September 2025

\bibitem{gwtc5}
{LVK Collaboration} 2026 Gwtc-5.0: An introduction to version 5.0 of the
  gravitational-wave transient catalog (\textit{Preprint} \eprint{2605.27223})
  \urlprefix\url{https://arxiv.org/abs/2605.27223}

\bibitem{hawking}
{LVK Collaboration} (LIGO Scientific, Virgo, and KAGRA Collaborations) 2025
  {\em Phys. Rev. Lett.\/} {\bf 135}(11) 111403
  \urlprefix\url{https://link.aps.org/doi/10.1103/kw5g-d732}

\bibitem{exploring}
{LVK Collaboration} 2025 {\em The Astrophysical Journal Letters\/} {\bf 993}
  L21 \urlprefix\url{https://doi.org/10.3847/2041-8213/ae0d54}

\bibitem{aVirgo}
{Virgo Collaboration} 2015 {\em Classical Quant. Grav.\/} {\bf 32} 024001
  \urlprefix\url{https://doi.org/10.1088/0264-9381/32/2/024001}

\bibitem{aLIGO}
{LIGO Scientific Collaboration} 2015 {\em Classical Quant. Grav.\/} {\bf 32}
  074001
  \urlprefix\url{https://iopscience.iop.org/article/10.1088/0264-9381/32/7/074001}

\bibitem{kagra}
{{KAGRA} Collaboration} 2019 {\em Nat. Astron.\/} {\bf 3} 35--40
  \urlprefix\url{https://doi.org/10.1038/s41550-018-0658-y}

\bibitem{superatt}
Accadia T {\em et~al.\/} 2011 {\em Journal of Low Frequency Noise, Vibration
  and Active Control\/} {\bf 30} 63--79 (\textit{Preprint}
  \eprint{https://doi.org/10.1260/0263-0923.30.1.63})
  \urlprefix\url{https://doi.org/10.1260/0263-0923.30.1.63}

\bibitem{coating6}
Reid S and Martin I~W 2016 {\em Coatings\/} {\bf 6} 61

\bibitem{Maggiore_2020}
Maggiore {\em et~al.\/} 2020 {\em JCAP\/} {\bf 2020} 050
  \urlprefix\url{https://dx.doi.org/10.1088/1475-7516/2020/03/050}

\bibitem{coba}
Branchesi {\em et~al.\/} 2023 {\em JCAP\/} {\bf 2023} 068
  \urlprefix\url{https://dx.doi.org/10.1088/1475-7516/2023/07/068}

\bibitem{CE1}
Reitze {\em et~al.\/} 2019 {\em arxiv.org/abs/1907.04833\/}
  \urlprefix\url{https://arxiv.org/abs/1907.04833}

\bibitem{CE2}
Evans {\em et~al.\/} 2021 {\em arxiv.org/abs/2109.09882\/}
  \urlprefix\url{https://arxiv.org/abs/2109.09882}

\bibitem{CE3}
{Hall} {\em et~al.\/} 2022 {\em Galaxies\/} {\bf 10} ISSN 2075-4434
  \urlprefix\url{https://www.mdpi.com/2075-4434/10/4/90}

\bibitem{et}
Abac A, Abramo R, Albanesi S {\em et~al.\/} 2025 The science of the {E}instein
  {T}elescope (\textit{Preprint} \eprint{arXiv.2503.12263})
  \urlprefix\url{https://arxiv.org/abs/2503.12263}

\bibitem{ET2010}
{ET Science Team} 2010 {\em Classical Quant. Grav.\/} {\bf 27} 194002
  \urlprefix\url{https://iopscience.iop.org/article/10.1088/0264-9381/27/19/194002}

\bibitem{ET2011}
{ET Science Team} 2011 {\em ET-0106C-10\/}
  \urlprefix\url{http://et-gw.eu/etdsdocument}

\bibitem{ET2020}
{ET Science Team} 2020 {\em ET-0028A-20\/}
  \urlprefix\url{http://et-gw.eu/etdsdocument}

\bibitem{o4bdatarelease}
{LVK Collaboration} 2026 Open data from ligo, virgo, and kagra through the
  second part of the fourth observing run (\textit{Preprint}
  \eprint{2605.27090}) \urlprefix\url{https://arxiv.org/abs/2605.27090}

\bibitem{etsenses}
Korobko M 2026 {ET} {S}ensitivity {C}urves
  \urlprefix\url{https://gitlab.et-gw.eu/pub/et-sensitivity-curves}

\bibitem{cesenses}
Kuns {\em et~al.\/} 2024 {CE-T}2000017-v9: Cosmic explorer strain and
  displacement sensitivity
  \urlprefix\url{https://dcc.cosmicexplorer.org/cgi-bin/DocDB/ShowDocument?docid=T2000017}

\bibitem{squeezed}
McCuller L {\em et~al.\/} 2020 {\em Phys. Rev. Lett.\/} {\bf 124}(17) 171102
  \urlprefix\url{https://link.aps.org/doi/10.1103/PhysRevLett.124.171102}

\bibitem{squeezing1}
Nishino {\em et~al.\/} 2024 {\em Phys. Rev. A\/} {\bf 110}(2) 022601
  \urlprefix\url{https://link.aps.org/doi/10.1103/PhysRevA.110.022601}

\bibitem{squeezing2}
Kimble {\em et~al.\/} 2001 {\em Phys. Rev. D\/} {\bf 65}(2) 022002
  \urlprefix\url{https://link.aps.org/doi/10.1103/PhysRevD.65.022002}

\bibitem{squeezing3}
Caves 1981 {\em Phys. Rev. D\/} {\bf 23}(8) 1693--1708
  \urlprefix\url{https://link.aps.org/doi/10.1103/PhysRevD.23.1693}

\bibitem{Iacovelli_2024}
Iacovelli {\em et~al.\/} 2024 {\em Journal of Cosmology and Astroparticle
  Physics\/} {\bf 2024} 085
  \urlprefix\url{https://dx.doi.org/10.1088/1475-7516/2024/10/085}

\bibitem{Grohmann2021}
Grohmann S 2021 Cryogenic developments towards the einstein telescope (et)
  Vortrag gehalten auf European Cryogenic Days (2021), Online, 3.--4. November
  2021 54.11.11; LK 01

\bibitem{koroveshi}
Koroveshi X {\em et~al.\/} 2023 {\em Phys. Rev. D\/} {\bf 108}(12) 123009
  \urlprefix\url{https://link.aps.org/doi/10.1103/PhysRevD.108.123009}

\bibitem{dupletsa2026}
Dupletsa U {\em et~al.\/} 2026 Assessing the impact of instrumental
  requirements on the scientific performance of the einstein telescope
  (\textit{Preprint} \eprint{2607.27311})
  \urlprefix\url{https://arxiv.org/abs/2607.27311}

\bibitem{coating1}
Craig K {\em et~al.\/} 2019 {\em Physical Review Letters\/} {\bf 122} 231102

\bibitem{coating4}
Granata M {\em et~al.\/} 2020 {\em Applied Optics\/} {\bf 59} A229--A235

\bibitem{coating5}
Cole G~D, Zhang W, Martin M~J, Ye J and Aspelmeyer M 2013 {\em Nature
  Photonics\/} {\bf 7} 644--650

\bibitem{Puppo2011}
Puppo and Ricci 2011 {\em General Relativity and Gravitation\/} {\bf 43}
  657--669

\bibitem{amann}
Amann {\em et~al.\/} 2020 {\em Rev. Sci. Instr.\/} {\bf 91} 094504
  \urlprefix\url{https://doi.org/10.1063/5.0018414}

\bibitem{harms}
Harms 2019 {\em Liv. Rev. Rel.\/} {\bf 22}
  \urlprefix\url{https://doi.org/10.1007/s41114-019-0022-2}

\bibitem{hutt}
Hutt {\em et~al.\/} 2017 {\em Bulletin of the Seismological Society of
  America\/} {\bf 107} 1402--1412 ISSN 0037-1106 (\textit{Preprint}
  \eprint{https://pubs.geoscienceworld.org/ssa/bssa/article-pdf/107/3/1402/2641899/BSSA-2016187.1.pdf})
  \urlprefix\url{https://doi.org/10.1785/0120160187}

\bibitem{mariusz}
Alhassan W {\em et~al.\/} 2026 Infrasound newtonian noise estimation at the
  einstein telescope candidate site sos enattos (\textit{Preprint}
  \eprint{2607.21154}) \urlprefix\url{https://arxiv.org/abs/2607.21154}

\bibitem{potsdam}
Bormann P {\em et~al.\/} 2009 {\em Seismic Wave Propagation and Earth models\/}
  (Deutsches GeoForschungsZentrum) pp 1--70
  \urlprefix\url{http://dx.doi.org/10.1142/9789811220944_0007}

\bibitem{CE4}
Evans {\em et~al.\/} 2023 {\em Technical Report CE–P2300018–v3\/}

\bibitem{nsfreport}
Kalogera {\em et~al.\/} 2024 Report from the {NSF} {MPS} {AC} subcommittee on
  next-generation gravitational-wave detector concepts
  \urlprefix\url{https://nsf-gov-resources.nsf.gov/files/mpsac-nggw-subcommittee-repor-2024-03-23-r.pdf?VersionId=uhdtblJUPYsavII5EUuSNQJcFdRr92q_}

\bibitem{iacovelli26}
Iacovelli F {\em et~al.\/} 2026 Not too close! evaluating the impact of the
  baseline on the localization of binary black holes by next-generation
  gravitational-wave detectors (\textit{Preprint} \eprint{2604.11871})
  \urlprefix\url{https://arxiv.org/abs/2604.11871}

\bibitem{naticchionietal2014}
Naticchioni {\em et~al.\/} 2014 {\em Classical Quant. Grav.\/} {\bf 31}
  \urlprefix\url{doi:10.1088/0264-9381/31/10/105016}

\bibitem{naticchionietal2020}
Naticchioni {\em et~al.\/} 2020 {\em J. Phys.: Conf. Ser.\/} {\bf 1468}
  \urlprefix\url{doi:10.1088/1742-6596/1468/1/012242}

\bibitem{digiovannietal2021}
{Di Giovanni} {\em et~al.\/} 2021 {\em Seismol. Res. Lett.\/} {\bf 92} 352--364
  \urlprefix\url{https://doi.org/10.1785/0220200186}

\bibitem{digiovanni2023}
{Di Giovanni} {\em et~al.\/} 2023 {\em Geophysical Journal International\/}
  {\bf 234} 1943--1964 (\textit{Preprint}
  \eprint{https://academic.oup.com/gji/article-pdf/234/3/1943/50285042/ggad178.pdf})
  \urlprefix\url{https://doi.org/10.1093/gji/ggad178}

\bibitem{saccorotti23}
Saccorotti {\em et~al.\/} 2023 {\em Eur. Phys. J. Plus\/} {\bf 138} 793
  \urlprefix\url{https://doi.org/10.1140/epjp/s13360-023-04395-2}

\bibitem{diaferia25}
Diaferia G {\em et~al.\/} 2025 {\em Solid Earth\/} {\bf 16} 441--456
  \urlprefix\url{https://se.copernicus.org/articles/16/441/2025/}

\bibitem{Diaferia26}
Diaferia G {\em et~al.\/} 2026 {\em Seismica\/} {\bf 5}
  \urlprefix\url{https://seismica.library.mcgill.ca/article/view/1809}

\bibitem{Bader_2022}
Bader {\em et~al.\/} 2022 {\em Classical and Quantum Gravity\/} {\bf 39} 025009
  \urlprefix\url{https://dx.doi.org/10.1088/1361-6382/ac1be4}

\bibitem{koley2022surface}
Koley {\em et~al.\/} 2022 {\em Class. Quant. Grav.\/} {\bf 39} 025008
  \urlprefix\url{https://iopscience.iop.org/article/10.1088/1361-6382/ac2b08}

\bibitem{alloccaetal2021}
Allocca {\em et~al.\/} 2021 {\em Eur. Phys. J. Plus\/}  511
  \urlprefix\url{https://doi.org/10.1140/epjp/s13360-021-01450-8}

\bibitem{janssens2022}
Janssens K {\em et~al.\/} 2022 {\em Phys. Rev. D\/} {\bf 106}(4) 042008
  \urlprefix\url{https://link.aps.org/doi/10.1103/PhysRevD.106.042008}

\bibitem{janssens2024}
Janssens K {\em et~al.\/} 2024 {\em PRD\/} {\bf 137} 102002
  \urlprefix\url{https://doi.org/10.1103/PhysRevD.109.102002}

\bibitem{digiovanni2025}
{Di Giovanni} {\em et~al.\/} 2025 {\em Class Quant Grav\/} {\bf 42}

\bibitem{pygwinc_ascl}
Rollins J~G {\em et~al.\/} 2020 pygwinc: Gravitational wave interferometer
  noise calculator Astrophysics Source Code Library, record ascl:2007.020
  \urlprefix\url{https://ascl.net/2007.020}

\bibitem{saulson94}
Saulson 1994 {\em Fundamentals of interferometric gravitational wave
  detectors\/} (London: World Scientific Publishing)

\bibitem{radpress}
Acernese F {\em et~al.\/} (The Virgo Collaboration) 2020 {\em Phys. Rev.
  Lett.\/} {\bf 125}(13) 131101
  \urlprefix\url{https://link.aps.org/doi/10.1103/PhysRevLett.125.131101}

\bibitem{teoF&D}
{Callen} H~B and {Welton} T~A 1951 {\em Physical Review\/} {\bf 83} 34--40

\bibitem{kagracryo1}
Ushiba {\em et~al.\/} 2021 {\em Classical and Quantum Gravity\/} {\bf 38}
  085013

\bibitem{kagracryo2}
others K 2016 {\em Journal of Physics: Conference Series\/} {\bf 716} 012017

\bibitem{sinica}
Guo {\em et~al.\/} 2024 {\em Acta Physica Sinica\/} {\bf 73} 050401

\bibitem{Longo2026}
Longo {\em et~al.\/} 2026 {\em Classical and Quantum Gravity\/} {\bf 43} 145002

\bibitem{difronzo2026}
Fronzo D and Liu 2026 Addressing rotational motion on gravitational waves
  detectors (\textit{Preprint} \eprint{2607.26430})
  \urlprefix\url{https://arxiv.org/abs/2607.26430}

\bibitem{tilt1}
Venkateswara {\em et~al.\/} 2014 {\em Review of Scientific Instruments\/} {\bf
  85} 015005

\bibitem{tilt2}
Coughlin {\em et~al.\/} 2018 {\em Phys. Rev. Lett.\/} {\bf 121}(22) 221104
  \urlprefix\url{https://link.aps.org/doi/10.1103/PhysRevLett.121.221104}

\bibitem{tilt3}
Calloni and othersn 2021 {\em The European Physical Journal Plus\/} {\bf 136}
  335

\bibitem{tilt4}
Allocca {\em et~al.\/} 2021 {\em The European Physical Journal Plus\/} {\bf
  136} 1069

\bibitem{peterson1993observations}
Peterson 1993 Observations and modeling of seismic background noise Tech. rep.
  US Geological Survey \urlprefix\url{https://doi.org/10.3133/ofr93322}

\bibitem{cvse}
{Virgo Collaboration} 2011 {\em Class. Quant. Grav.\/} {\bf 29}(2)
  \urlprefix\url{https://iopscience.iop.org/article/10.1088/0264-9381/29/2/025005/pdf}

\bibitem{Fiori2020}
Fiori {\em et~al.\/} 2020 Environmental noise in gravitational-wave
  interferometers {\em Handbook of Gravitational Wave Astronomy\/} (Singapore:
  Springer Singapore) pp 1--72
  \urlprefix\url{https://doi.org/10.1007/978-981-15-4702-7_10-1}

\bibitem{o3noise}
{Virgo Collaboration} 2022 {\em Classical and Quantum Gravity\/} {\bf 39}
  235009 \urlprefix\url{https://dx.doi.org/10.1088/1361-6382/ac776a}

\bibitem{accadia2010noise}
{Virgo Collaboration} 2010 {\em Classical and Quantum Gravity\/} {\bf 27}
  194011
  \urlprefix\url{https://iopscience.iop.org/article/10.1088/0264-9381/27/19/194011}

\bibitem{bonnefoy2006nature}
Bonnefoy-Claudet {\em et~al.\/} 2006 {\em Earth Sci. Rev.\/} {\bf 79} 205--227

\bibitem{hoshino}
Hoshino S {\em et~al.\/} 2024 {\em Progress of Theoretical and Experimental
  Physics\/} {\bf 2024} 103F01 ISSN 2050-3911 (\textit{Preprint}
  \eprint{https://academic.oup.com/ptep/article-pdf/2024/10/103F01/60221269/ptae108.pdf})
  \urlprefix\url{https://doi.org/10.1093/ptep/ptae108}

\bibitem{saccorotti}
Saccorotti {\em et~al.\/} 2011 {\em Bull. Seis. Soc. Am.\/} {\bf 101}
  \urlprefix\url{https://doi.org/10.1785/0120100203}

\bibitem{ligowind}
Schofield R 2002 Ligo - t020104-x0: Seismic measurements at the stateline wind
  project - and a prediction of the seismic signal that the proposed maiden
  wind project would produce at ligo
  \urlprefix\url{https://dcc-lho.ligo.org/public/0027/T020104/000/T020104-00.pdf}

\bibitem{ligosusp}
Matichard F {\em et~al.\/} 2015 {\em Classical and Quantum Gravity\/} {\bf 32}
  185003

\bibitem{Ballardin2001VirgoSuperattenuator}
Ballardin G {\em et~al.\/} 2001 {\em Review of Scientific Instruments\/} {\bf
  72} 3643--3652

\bibitem{kagrasusp}
Ushiba T 2021 {\em Classical and Quantum Gravity\/} {\bf 38} 085013

\bibitem{invertedP}
Losurdo G {\em et~al.\/} 1999 {\em Review of Scientific Instruments\/} {\bf 70}
  2507--2515 ISSN 0034-6748 \urlprefix\url{https://doi.org/10.1063/1.1149783}

\bibitem{infra}
Posmentier E~S 1974 {\em Journal of Geophysical Research (1896-1977)\/} {\bf
  79} 1755--1760 (\textit{Preprint}
  \eprint{https://agupubs.onlinelibrary.wiley.com/doi/pdf/10.1029/JC079i012p01755})
  \urlprefix\url{https://agupubs.onlinelibrary.wiley.com/doi/abs/10.1029/JC079i012p01755}

\bibitem{Weiss:1972}
Weiss 1972 Electromagnetically coupled broadband gravitational antenna
  Quarterly Progress Report RLE QPR 105 MIT Research Laboratory of Electronics
  \urlprefix\url{https://dcc.ligo.org/LIGO-MIT-ME-1973001/public}

\bibitem{beccaria1998relevance}
Beccaria {\em et~al.\/} 1998 {\em Classical Quant. Grav.\/} {\bf 15} 3339
  \urlprefix\url{https://iopscience.iop.org/article/10.1088/0264-9381/15/11/004}

\bibitem{Hughes1998SeismicGravityGradient}
Hughes S~A and Thorne K~S 1998 {\em Physical Review D\/} {\bf 58} 122002

\bibitem{cella}
Cella G 2000 Off-line subtraction of seismic newtonian noise {\em Recent
  Developments in General Relativity\/} ed Casciaro B, Fortunato D,
  Francaviglia M and Masiello A (Milano: Springer Milan) pp 495--503 ISBN
  978-88-470-2113-6

\bibitem{harms15}
Harms J and Paik H~J 2015 {\em Phys. Rev. D\/} {\bf 92}(2) 022001
  \urlprefix\url{https://link.aps.org/doi/10.1103/PhysRevD.92.022001}

\bibitem{Coughlin_2016}
Coughlin M, Mukund N, Harms J, Driggers J, Adhikari R and Mitra S 2016 {\em
  Classical and Quantum Gravity\/} {\bf 33} 244001
  \urlprefix\url{https://doi.org/10.1088/0264-9381/33/24/244001}

\bibitem{badaracco2019}
Badaracco {\em et~al.\/} 2019 {\em Class. Quantum Grav.\/} {\bf 36}
  \urlprefix\url{https://doi.org/10.1088/1361-6382/ab28c1}

\bibitem{Badaracco2020}
Badaracco {\em et~al.\/} 2020 {\em Classical and Quantum Gravity\/} {\bf 37}
  195016 \urlprefix\url{https://dx.doi.org/10.1088/1361-6382/abab64}

\bibitem{Koley2024}
Koley {\em et~al.\/} 2024 {\em Phys. Rev. D\/} {\bf 110}(2) 022002
  \urlprefix\url{https://link.aps.org/doi/10.1103/PhysRevD.110.022002}

\bibitem{Creighton_2008}
Creighton T 2008 {\em Classical and Quantum Gravity\/} {\bf 25} 125011
  \urlprefix\url{https://doi.org/10.1088/0264-9381/25/12/125011}

\bibitem{Badaracco:2021prd}
Badaracco F {\em et~al.\/} 2021 {\em Physical Review D\/} {\bf 104}(4) 042006
  \urlprefix\url{https://link.aps.org/doi/10.1103/PhysRevD.104.042006}

\bibitem{harms2022}
Harms {\em et~al.\/} 2022 {\em The European Physical Journal Plus\/} {\bf 137}
  687 ISSN 2190-5444
  \urlprefix\url{https://doi.org/10.1140/epjp/s13360-022-02851-z}

\bibitem{schillings2026numericalframeworknewtoniannoiseestimation}
Schillings P {\em et~al.\/} 2026 A numerical framework for newtonian-noise
  estimation at the einstein telescope: 2-d simulations beyond the plane-wave
  approximation (\textit{Preprint} \eprint{2603.15424})
  \urlprefix\url{https://arxiv.org/abs/2603.15424}

\bibitem{Fiori_2026}
Fiori I {\em et~al.\/} 2026 {\em Classical and Quantum Gravity\/} {\bf 43}
  115003 \urlprefix\url{https://doi.org/10.1088/1361-6382/ae68d2}

\bibitem{acernese2004properties}
Acernese {\em et~al.\/} 2004 {\em Classical and Quantum Gravity\/} {\bf 21}
  S433
  \urlprefix\url{https://iopscience.iop.org/article/10.1088/0264-9381/21/5/008/pdf}

\bibitem{virgo2006}
{Virgo Collaboration} 2006 {\em J. Phys. Conf. Ser.\/} {\bf 32} 80
  \urlprefix\url{https://iopscience.iop.org/article/10.1088/1742-6596/32/1/013}

\bibitem{poli}
{Poli} {\em et~al.\/} 2020 {\em Sci. Rep.\/} {\bf 10} 9404
  \urlprefix\url{https://doi.org/10.1038/s41598-020-66368-0}

\bibitem{piccinini}
{Piccinini} {\em et~al.\/} 2020 {\em Sci. Rep.\/} {\bf 10} 16487
  \urlprefix\url{https://doi.org/10.1038/s41598-020-73102-3}

\bibitem{cirone}
Cirone A {\em et~al.\/} 2018 {\em Review of Scientific Instruments\/} {\bf 89}
  114501

\bibitem{Yokozawa:2023icrc}
Yokozawa T (KAGRA) 2024 {Investigation of the environmental noise at KAGRA
  detector} {\em {Proceedings of 38th International Cosmic Ray Conference ---
  PoS(ICRC2023)}\/} ({\em Proceedings of Science\/} vol 444) p 1557
  \urlprefix\url{https://pos.sissa.it/444/1557/}

\bibitem{boxberg}
Boxberg M~e~a 2026 Reducing the impact of wind turbine vibrations on the
  einstein telescope (poster at the {XVI ET} symposium)
  \urlprefix\url{https://indico.ego-gw.it/event/960/contributions/10211/attachments/6023/10827/Poster%20ETS%20Aachen.pdf}

\bibitem{wood}
Heiduschke A {\em et~al.\/} 2010 Tubular timber poles for small wind turbines
  {\em Proceedings of the 11th World Conference on Timber Engineering\/} vol~3
  (Univ. Acad. Pr.) pp 1821--1825

\bibitem{vanBeveren_2023}
{van Beveren} {\em et~al.\/} 2023 {\em Classical Quant. Grav.\/} {\bf 40}
  205008 \urlprefix\url{https://dx.doi.org/10.1088/1361-6382/acf3c8}

\bibitem{kelleter2026nnnnneuralnetworksnewtonian}
Kelleter J {\em et~al.\/} 2026 Nnnn: Neural networks for newtonian noise
  mitigation at the einstein telescope (\textit{Preprint} \eprint{2606.19907})
  \urlprefix\url{https://arxiv.org/abs/2606.19907}

\bibitem{Andric2020}
{Andric} T and {Harms} J 2020 {\em Journal of Geophysical Research (Solid
  Earth)\/} {\bf 125}(10) e2020JB020401
  \urlprefix\url{https://agupubs.onlinelibrary.wiley.com/doi/full/10.1029/2020JB020401}

\bibitem{Schillings_2025}
Schillings and Erdmann 2025 {\em Classical and Quantum Gravity\/} {\bf 42}
  065025 \urlprefix\url{https://doi.org/10.1088/1361-6382/adb898}

\bibitem{schillings2026b}
Schillings and Erdmann 2026 Optimization and robustness of cost-efficient
  seismic arrays for newtonian noise cancellation at the einstein telescope
  (\textit{Preprint} \eprint{2606.21110})
  \urlprefix\url{https://arxiv.org/abs/2606.21110}

\bibitem{Ophardt_2026}
Ophardt {\em et~al.\/} 2026 {\em Classical and Quantum Gravity\/} {\bf 43}
  125001 \urlprefix\url{https://doi.org/10.1088/1361-6382/ae76b7}

\bibitem{Negri:2026clm}
Negri L {\em et~al.\/} 2026 {Impact of the Einstein Telescope's duty cycle on
  the estimation of binary black holes parameters} (\textit{Preprint}
  \eprint{2606.19201})

\bibitem{mnvirgo}
Cirone A {\em et~al.\/} 2019 {\em Classical and Quantum Gravity\/} {\bf 36}
  225004

\bibitem{West1985}
West M 1985 {\em Identification and Protection of Existing and Potential
  Observatory Sites\/} (Dordrecht: Springer Netherlands) pp 707--712 ISBN
  978-94-009-5392-5
  \urlprefix\url{https://doi.org/10.1007/978-94-009-5392-5_39}

\bibitem{Goddi2025}
Goddi C {\em et~al.\/} 2025 {\em Astronomy \& Astrophysics\/} {\bf 699} A265

\bibitem{FCC11745_NRQZ_1958}
{Federal Communications Commission} 1958 {Docket No. 11745: Establishment of
  the National Radio Quiet Zone} Tech. rep. Federal Communications Commission
  (FCC) adopted November 19, 1958
  \urlprefix\url{https://www.gb.nrao.edu/nrqz/FCC_Docket_11745_NRQZ.pdf}

\bibitem{Borlaff2025_Megaconstellations}
Borlaff A~S, Marcum P~M and Howell S~B 2025 {\em Nature\/} {\bf 648} 51--57
  \urlprefix\url{https://www.nature.com/articles/s41586-025-09759-5}

\end{thebibliography}
\end{document}